\documentclass[aps,prd,12pt,nofootinbib,superscriptaddress]{revtex4}
\usepackage{epsfig}
\usepackage{graphicx}
\usepackage{cancel}
\usepackage{amsmath}
\usepackage{amssymb}
\usepackage{mathrsfs}
\usepackage{verbatim}
\usepackage{dutchcal}
\usepackage{comment}
\usepackage{float}
\usepackage[normalem]{ulem}
\usepackage{xcolor}
\usepackage{slashed}

\newcounter{fig}   

\newcommand{\Tr}{{\rm Tr}}

\newcommand{\bea}{\begin{eqnarray}}
\newcommand{\eea}{\end{eqnarray}}
\newcommand{\be}{\begin{equation}}
\newcommand{\ee}{\end{equation}}

\def\({\left(}
\def\){\right)}

\newcommand{\re}[1]{(\ref{#1})}

\def\rlx{\relax\leavevmode}
\def\IR{\rlx\hbox{\rm I\kern-.18em R}}
\def\one{\hbox{{1}\kern-.25em\hbox{l}}}

\newcommand{\eqn}{\begin{eqnarray}}
\newcommand{\eqnx}{\end{eqnarray}}

\begin{document}

\title{Isospinning ${\mathbb C}P^2$ solitons}

\author{
Yuki Amari
}
\affiliation{Department of Physics $\&$ Research and Education Center for Natural Sciences, Keio University, Hiyoshi 4-1-1, Yokohama, Kanagawa 223-8521, Japan}
\author{Sergei Antsipovich}
\affiliation{
Department of Theoretical Physics and Astrophysics,
Belarusian State University, Minsk 220004, Belarus
}
\author{Muneto Nitta}
\affiliation{Department of Physics $\&$ Research and Education Center for Natural Sciences, Keio University, Hiyoshi 4-1-1, Yokohama, Kanagawa 223-8521, Japan}
\affiliation{
International Institute for Sustainability with Knotted Chiral Meta Matter(SKCM$^2$), Hiroshima University, 1-3-2 Kagamiyama, Higashi-Hiroshima, Hiroshima 739-8511, Japan
}
\author{Yakov Shnir}
\affiliation{BLTP, JINR, Dubna 141980, Moscow Region, Russia}
\affiliation{Institute of Physics,
Carl von Ossietzky University Oldenburg, Germany
Oldenburg D-26111, Germany}
\affiliation{Hanse-Wissenschaftskolleg, Lehmkuhlenbusch 4, 27733 Delmenhorst, Germany}

\date{\today}

\begin{abstract}
We study stationary rotating topological solitons  in (2+1)-dimensional  ${\mathbb C}P^2$ non-linear sigma model with a stabilizing potential term. We find families of $U(1)\times U(1)$ symmetric solutions with topological degrees larger than 2, which
 have two angular frequencies
and
are labelled by two
(one topological and the other non-topological)
winding numbers  $k_1>k_2$.
We discuss properties of these solitons and investigate the domains of their existence.
\end{abstract}


\maketitle
\section{Introduction}
Topological and non-topological solitons caused a lot of attraction over the last decades.
These particle-like regular field configurations naturally appearing in a wide variety of
physical systems, for a review see, for example Refs.~\cite{Manton:2004tk,Shnir:2018yzp}.
Famous examples of topological solitons in (3+1)-dimensions are monopoles in the
Yang-Mills-Higgs theory \cite{tHooft:1974kcl,Polyakov:1974ek},
Skyrmions \cite{Skyrme:1961vq,Skyrme:1962vh}, and Hopfions \cite{Faddeev:1975tz,Faddeev:1996zj}.
Notably, spectrum of perturbative excitations of such localized solutions possess
translational, rotational and isorotational zero modes.
In the Skyrme model, the rotational degrees of freedom of a soliton account, after quantization,
for the spin and isospin quantum numbers of a corresponding nucleon. The traditional approach to study the
spinning Skyrmions is related with the rigid body approximation \cite{Adkins:1983ya,Adkins:1983hy},
in which
an assumption is made that a soliton rotates without changing its shape but its spatial
and internal orientations  have harmonic time dependence. Similar approximation was used in early
study of isospinning solutions in (2+1)-dimensional baby Skyrme model \cite{Piette:1994mh}.
The self-consistent study of isospinning solitons beyond rigid body approximation is a significant
computational challenge; Only recently  such internally rotating configurations were constructed in
the baby Skyrme model in 2+1 dimensions \cite{Halavanau:2013vsa,Battye:2013tka}
 and
the Faddeev-Skyrme model \cite{Battye:2013xf,Harland:2013uk},
and Skyrme model \cite{Battye:2014qva}
in 3+1 dimensions.

Several interesting general observations can be made for classically isospinning solitons.
Firstly,
a potential term must be included in the corresponding Lagrangians in order to stabilize the configuration.
This term gives a mass to linearized excitations, and it also yields an upper bound on the allowed range of values
of frequency of internal rotations. Secondly, a harmonic time dependency of such configurations allows to circumvent
the Derrick's no-go theorem for finite energy localized field configurations \cite{Derrick:1964ww}.
Such a time dependence
stabilizes
not only nontopological
solitons such as Q-balls \cite{Coleman:1985ki,Shnir:2018yzp}
 but also stabilize unstable topological solitons.
For instance, as an example of the latter,
the $O(3)$ sigma model with a potential term does not admit
stable topological lump solitons since they are unstable to shrink.
Nevertheless, it admits
time-dependent topological lump solitons called Q-lumps
\cite{Leese:1991hr,Abraham:1991ki,Ward:2003un}.
In addition, topologically stable
solitons can be further turned to
 Q-topological solitons,
such as
a Q-kink
\cite{Abraham:1992vb,Abraham:1992qv} and
Q-lump strings ending on a domain wall
\cite{Gauntlett:2000de,Isozumi:2004vg,Eto:2008mf}
in the $O(3)$ model with an easy axis potential.
Thirdly, it was observed that the angular momentum of isospinning solitons  depends linearly on the
Noether charge of the configuration \cite{Radu:2008pp}. Finally, the isorotations can break the symmetries of the
multisoliton configuration and split it into components, as the angular frequency increases above a certain limit \cite{Halavanau:2013vsa,Battye:2013tka,Battye:2014qva}.

 The $O(3)$ non-linear sigma model is equivalent to the
${\mathbb C}P^1$ model \cite{Leese:1989gi,Sutcliffe:1991aua}.
Further generalization of this construction
leads to the (2+1)-dimensional ${\mathbb C}P^{N}$ non-linear sigma model \cite{Golo:1978de,DAdda:1978vbw,Din:1980jg} and
${\mathbb C}P^N$ Faddeev-Skyrme model \cite{Ferreira:2010jb,Amari:2015sva}.
The ${\mathbb C}P^2$ non-linear sigma model has attracted
attentions recently in condensed matter physics as well: it
can be obtained as a continuum limit of the $SU(3)$ ferromagnetic Heisenberg model \cite{Ivanov,Smerald}, and modifications
of this model were studied lately
with  a Dzyaloshinskii-Moriya interaction term \cite{Akagi:2021dpk,Amari:2022boe},
with competing easy-axis ferromagnetic and antiferromagnetic interactions \cite{Zhang:2022lyz} or with a Skyrme-like term
\cite{Akagi:2021lva}.
As for Q-solitons,
the ${\mathbb C}P^N$ model (and Grassmannian generalization)
admit parallel Q-kinks
\cite{Gauntlett:2000ib,Tong:2002hi,Eto:2008dm},
Q-lump strings stretched between parallel domain walls
\cite{Isozumi:2004vg,Eto:2008mf,
}
and Q-domain wall networks \cite{Eto:2007uc}.
Nontopological Q-solitons in the ${\mathbb C}P^N$ model were also investigated in
Refs.~\cite{Klimas:2017eft,Klimas:2018ywv,Klimas:2022ghu}.

In this paper, we investigate  classically stable isospining multiple topological soliton solutions of the ${\mathbb C}P^2$ non-linear sigma model with a stabilizing potential. We construct  solutions of topological degrees from 2 to 5 numerically and  perform the analysis of their critical behavior.

This paper is organized as follows.
In Sec.~\ref{sec:model}, we introduce the model and define the Ansatz and the boundary
conditions for the isospinning topolocally non-trivial ${\mathbb C}P^2$ solitons.
The numerical results are presented in
Sec.~\ref{sec:results} where we also considered limiting behavior of the solutions.
We give our conclusions and remarks in the final section \ref{sec:summary}.

\section{The model}\label{sec:model}

\subsection{Action}
In this paper, we consider a $2+1$
dimensional nonlinear sigma model with the target space $\mathbb{C}P^{N-1} \backsimeq
SU(N)/[SU(N-1)\times U(1)]$, i.e., $\mathbb{C}P^{N-1}$ nonlinear sigma model.
The model is defined by the Lagrangian \cite{Golo:1978de,DAdda:1978vbw,Din:1980jg,Zakrzewski}
\be \label{lag}
\cal L =\frac12 \Tr (\partial_\mu \mathbf{n}\, \partial^\mu \mathbf{n}) - V(\mathbf{n})
\, ,
\ee
where $\mathbf{n}= n^a\lambda_a$, $(a=1,2\dots N^2-1)$ and $\lambda_a$ are the $SU(N)$ generators
subject to normalization $\Tr (\lambda_a \lambda_b) = 2\delta_{ab}$.
The fields $n^a$ satisfy the $N^2$ constraints:
\be
n^a n^a =\frac{2(N-1)}{N}\, , \qquad n^a=\frac{N}{2(N-2)} d_{abc} n^b n^c\,
\ee
where the totally symmetric third rank tensor  $d_{abc}=\frac14 \Tr(\lambda_a\{\lambda_b,\lambda_c\})$ yield the cubic
Casimir operator, $d_{abc}n^a n^b n^c = 4(N-1)(N-2)/N^2$.
Some of the constraints are redundant, and the actual number of degrees of freedom the fields possess is $\dim[\mathbb{C}P^{N-1}]=2(N-1)$.
The first term in Eq.~\re{lag} is quadratic in derivatives, which is the usual Lagrangian of the
non-linear sigma-model. Clearly, it is invariant with respect to the global transformation of the
field $\mathbf{n} \to U^\dagger  \mathbf{n} U$, where $U\in ~SU(N)$. When the potential is absent, the symmetry is spontaneously broken to $SU(N-1)\times U(1)$ of which generators commute with a vacuum configuration. This symmetry can explicitly be broken by
the potential term $V(\mathbf{n})$ which is introduced to stabilize the configuration.
We will consider a potential which will break this symmetry to $H=U(1)^N$, i.e.
to the diagonal subgroup of $U(N)$. 

Finite energy solutions of the model \re{lag} require that the field $\mathbf{n}$ has to approach a vacuum value on the spacial
boundary. This corresponds to a one-point compactification of the domain space $\mathbb{R}^2 \to S^2$. Concequently, the field is defined as a map
$\mathbf{n}: S^2 \to \mathbb{C}P^{N-1}$ characterized by the topological invariant $Q_{\rm top} \in \pi_2(\mathbb{C}P^{N-1}) =
\mathbb{Z}$. Explicitly, it is defined by the integral
\be
Q_{\rm top}=\frac{1}{8\pi}\int d^2x \varepsilon^{ij} f_{abc} n^a \partial_i n^b \partial_j n^c
\label{topcharge}
\ee
with the $SU(N)$ structure constants $f_{abc}= \frac{i}{4}\Tr(\lambda_a[\lambda_b,\lambda_c])$.

The parametrization of the $\mathbb{C}P^{N-1}$ model in terms of the field $\mathbf{n}$  is not optimal for constructing soliton solutions because it involves
an additional set of complicated non-linear constraints, although it is convenient for the model building, especially, introducing a potential term that supports soliton solutions. Instead, a set of homogeneous coordinates on $\mathbb{C}P^{N-1}$, $Z^\alpha~(\alpha=1,2\dots N)$,
can be introduced. The complex $N$-componentvector $Z=(Z^1, Z^2,..., Z^N)^\tau$ satisfies $ Z^\dagger Z =1 $, where $\tau$ denotes the transposition operator and the dagger indicates Hermitian conjugation.
The components of this vector are related to the field $\mathbf{n}$ as
$$
n^a = Z^\dagger \lambda_a Z \, .
$$
In terms of $Z$ the Lagrangian \re{lag} becomes
\be
{\cal L}= 4 (D_\mu Z)^\dagger (D^\mu Z) - V(|Z|)
\label{lagZ}
\ee
while the topological charge \re{topcharge} is given by
\be
Q_{\rm top}=-\frac{i}{2\pi}\int d^2 x \varepsilon^{ij} (D_i Z)^\dagger (D_j Z)=
\int d^2 x \varepsilon^{ij} \partial_i A_j
\label{topZ}
\ee
where the covariant derivative is defined as
$D_\mu Z=\partial_\mu Z - i A_\mu Z  $ with the induced connection
$A_\mu=-i Z^\dagger \partial_\mu Z$. It follows from the definition that
\be
(D_\mu Z)^\dagger (D^\mu Z) = \partial_\mu Z^\dagger \cdot
\partial^\mu Z - (\partial_\mu Z^\dagger \cdot Z)\cdot
(Z^\dagger \cdot \partial^\mu  Z)\, .
\ee

The 
stress-energy tensor associated with the Lagrangian \eqref{lag} is
\be
T_{\mu\nu} = 4 \left\{ (D_\mu Z)^\dagger (D_\nu Z) + (D_\nu Z)^\dagger (D_\mu Z) \right\}- {\cal L} g_{\mu\nu}\, .
\label{energy-stress}
\ee
Thus, the energy is given by
\be
E=\int d^2 x T_{00} = \int d^2 x\left[4\left\{
(D_0 Z)^\dagger (D_0 Z) + (D_i Z)^\dagger (D_i Z)\right\} + V(|Z|)\right]
\label{T00}
\ee
and the angular momentum of the spinning configuration reads
\be
J=\int d^2 x T_{0\theta} =4 \int d^2 x \left[(D_0 Z)^\dagger (D_\theta Z)
+ (D_\theta Z)^\dagger (D_0 Z) \right] \, ,
\label{J}
\ee
where the
variables $(r,\theta)$ stand for the standard polar coordinates on the plane $\mathbb{R}^2$.
\subsection{Isospinning $\mathbb{C}P^{1}$ solitons}
The simplest possible case $N=2$ corresponds to the $\mathbb{C}P^{1}$ (or equivalently $O(3)$)
non-linear sigma model with a symmetry breaking
potential term \cite{Ward:2003un,Mareike}. Although in such a case the Derrick's theorem excludes the possibility of the
existence of a static soliton solution \cite{Derrick:1964ww}, the internal rotations of a localized
configuration appear as a mechanism that may stabilize it.

Considering a potential breaking the global $O(3)$ symmetry to $O(2)$, e.g., $V(\mathbf{n})=1-\left(n^3\right)^2$, one can make use
of the unbroken global $O(2)$ symmetry and consider stationary isorotations of the fields around $\vec{n}_\infty=(0,0,1)$,
\be
(n^1+in^2) \mapsto (n^1+in^2)e^{i\omega t}
\label{rot1}
\ee
where $\omega$ is an internal frequency. The corresponding Noether current is
$j_\mu=\varepsilon_{abc}n^a_\infty n^b \partial_\mu n^c$, and the conserved non-topological charge of the stationary isospinning configuration, the isospin \re{J}, is
\be
J=\omega\int d^2 x\{(n^1)^2 + (n^2)^2\} \equiv \omega \Upsilon
\label{Ncharge}
\ee
where $\Upsilon$ is the moment of inertia.

The total energy functional of the model of the type \re{lag} in $d$ spacial dimensions
includes three terms,
\begin{equation}
    E = \frac12 \int d^d x \Tr (\partial_\mu \mathbf{n}\, \partial^\mu \mathbf{n}) + \int d^d x V(\mathbf{n}) + \frac{J^2}{2 \Upsilon}
\equiv E_2 + E_0 + E_{\omega}\, .
\end{equation}
Under the scaling transformations $x\to \lambda x $, this functional transforms as $E\to \lambda^{2-d}E_2 + \lambda^{-d}E_0
+\lambda^{d+2}E_{\omega}$. Therefore, one obtains
\be
\frac{d E(\lambda)}{d \lambda} = (2-d)\lambda^{1-d}E_2 - d \lambda^{-d-1}E_0 + d \lambda^{d+1}E_{\omega} \, ,
\label{energ_terms}
\ee
and the stationary point of the total energy functional $E$ exists at $\lambda=1$ if
\be
\label{Derrik-Q}
(2-d)E_2 + d (E_{\omega}-E_0) = (2-d)E_2 + d \left(\frac{J^2}{2 \Upsilon} -E_0\right) = 0 \, .
\ee
Thus, in $d=2$ spacial dimensions, the potential term can be balanced by the rotational energy
providing the stability of a localized configuration \cite{Ward:2003un,Mareike}.
The corresponding virial relation is
\be
\label{virial}
E_0=E_{\omega}=\frac{J^2}{2 \Upsilon} \, .
\ee

Notably, these dynamically stabilized
Q-lumps carry two conserved charges, the topological charge $Q_{\rm top}$ \re{topcharge} and the Noether
charge $Q$ \re{Ncharge} \cite{Ward:2003un}. In terms of inhomogeneous coordinates on $\mathbb{C}P^{1}$,
 $u\equiv\bar{Z}^1/\bar{Z}^2=(n^1+in^2)/(1-n^3)$ with the bar standing for the complex conjugate
, the isorotations \re{rot1} correspond to the $U(1)$ transformation $u\mapsto ue^{i\alpha}$  with
stationary phase $\alpha$.

\subsection{Isospinning $\mathbb{C}P^{2}$ solitons}
Next, we extend the above results to the case of the $\mathbb{C}P^{2}$ model. We first consider $\mathbb{C}P^1$ embedding solitons in the $\mathbb{C}P^2$ nonlinear sigma model. After that, we proceed to a discussion for genuine $\mathbb{C}P^2$ isospinning solitons.

\subsubsection{Embedded $\mathbb{C}P^{1}$ solitons in the $\mathbb{C}P^{2}$ nonlinear sigma model}
\label{subsec:embeding}

 Considering $Z$ as the fundamental
field, which takes values in $S^5/S^1 \approx \mathbb{C}P^2$,
we can exploit the symmetry of the Lagrangian \re{lagZ} with respect to transformations $Z\to Z h$ where
$h \in H=U(1)\times U(1)$. In particular, one can consider the usual definition
of the $SU(3)$ electric charge generator $\widetilde Q =(\lambda_3 +\frac{\lambda_8}{\sqrt 3})/2$ which acts on the
field $Z$ as
\be
Z \to Z \Lambda, \quad \Lambda= e^{i \alpha \widetilde Q}.
\ee
The corresponding Noether current is
\begin{equation}
j_{\mu }^{(\widetilde Q)}=\frac{i}{2}\left
(\frac{\partial L }{\partial \left ( \partial_{\mu } Z \right )}\Lambda Z-Z^{\dagger }\Lambda
\frac{\partial L }{\partial \left ( \partial_{\mu } Z^{\dagger } \right )} \right )
\label{Q-Noether1}
\end{equation}
and, up to the global $U(1)$ transformations, one can parametrize the isorotating configuration as
\be
Z=\(\cos F(r)e^{i \alpha},~\sin F(r) \cos G(r),~\sin F(r) \sin G(r) \)^\tau\, ,
\label{Z1}
\ee
where the phase $\alpha = \omega t + k \theta$ with a positive integer $k$.

In terms of the field $\mathbf{n}$,
this parametrization corresponds to the stationary isorotations of the components, which lie in two $su(2)$ subalgebras of
$su(3)$, associated with corresponding  simple roots,
\be
(n^1+in^2) \mapsto (n^1+in^2)e^{i\omega t}, ~~ (n^4+in^5) \mapsto (n^4+in^5)e^{i\omega t}\, ,
\label{rotcomp-1}
\ee
while the components $(n^6,~n^7)$ and  $(n^3,~n^8)$, which lie in the third $su(2)$ subalgebra, associated with composite
simple root of $su(3)$ and its diagonal  subalgebra, respectively, are not rotating. Note that the
Ansatz \re{Z1} corresponds to $n^7=0$.

The choice of the potential term is crucial to secure existence of the stable isospinning solutions \cite{Ward:2003un}. Analogously to Q-balls, the  time-dependent
$\mathbb{C}P^2$ solitons are only stable within a certain frequency range which depends strongly on the choice of the
potential term. The effective potential should support localized configurations, the asymptotic form of the field equations yields the upper critical value of the angular frequencies.
On the other hand, the effective potential
should support a metastable vacuum, it gives the lower critical value of the  frequencies.

The vacuum boundary conditions imply that on the spacial boundary $S^1$ the potential is vanishing.
Naively, we can consider the $U(1)\times U(1)$ symmetric potential
\be
V= \frac{\mu^2}{4} \left[\frac43 - (n^3)^2 - (n^8)^2 \right]
\label{pot}
\ee
where $\mu$ is a constant. Since the Ansatz \re{Z1} gives $n^3=\cos^2 F - \sin^2 F \cos^2 G$ and
$n^8=
(\cos^2 F - \frac12 \sin^2 F + \frac32 \sin^2 F \cos(2G))/\sqrt{3}$,
one can impose the vacuum boundary condition
\be
F(\infty)=0, ~~~G(\infty)=0
\label{BC}
\ee
i.e., $n^3(\infty)=1, ~n^8(\infty)=1/\sqrt 3$. However, such a choice leads to a trivial
$\mathbb{C}P^{1}$ embedded solution.
Note that in the $\mathbb{C}P^1$ case, the double vacuum potential $V=1-(n^3)^2$
brings a strong attraction between solitons and consequently, stable multi-soliton
solutions possess the axial symmetry. Thus, it is reasonable to employ the axial symmetric
ansatz \eqref{Z1} in order to obtain stable soliton solutions with the potential \eqref{pot},
which is a generalization of the $\mathbb{C}P^1$ double vacuum potential.

\subsubsection{Isospinning $\mathbb{C}P^{2}$ solitons}

We are now in a position to study non-embedded isospinning
solitons in the $\mathbb{C}P^2$ nonlinear sigma model. In order to
construct non-embedded isospinning $\mathbb{C}P^{2}$ solitons, we
will use another rotationally-invariant Ansatz \be Z=\(\cos
F(r),~\sin F(r) \cos G(r)e^{i \varphi},~\sin F(r) \sin G(r)e^{i
\psi} \)^\tau\, , \label{Z} \ee where $\varphi=\omega_1 t + k_1
\theta$ and $\psi=\omega_2 t + k_2 \theta$ with angular
frequencies $\omega_{1,2}$ and integer winding numbers $k_{1,2}$.
The two corresponding  $U(1)$ generators with the left/right
action on $Z$ are
\be
t_\varphi = ~{\rm diag}~\{0,1,0\},\quad
t_\psi = ~{\rm diag}~\{0,0,1\}\, . \label{u1gens}
\ee
Indeed,
under rotational transformation $\theta \to \theta + \delta
\theta$, the field $Z$ varies as
\be
\begin{split}
Z &\to e^{i k_1 t_\varphi \delta \theta }e^{i k_2 t_\psi \delta \theta } Z =
e^{\frac{i}{3} (k_1+k_2) \delta \theta }
e^{-\frac{i k_1}{2} \lambda_3 \delta \theta }
e^{\frac{i}{2 \sqrt3} (k_1-2k_2)\lambda_8 \delta \theta }\,Z\\
&=e^{-\frac{i k_1}{2} \lambda_3 \delta \theta }
e^{\frac{i}{2 \sqrt3} (k_1-2k_2)\lambda_8 \delta \theta }\,Z ~~~({\rm mod~global~}U(1))\, .
\end{split}
\ee

In the field notations $\mathbf{n}$,
this parametrization corresponds to the stationary isorotations of the components
\be
(n^1+in^2) \mapsto (n^1+in^2)e^{i\omega_1 t}, ~~ (n^4+in^5) \mapsto (n^4+in^5)e^{i\omega_2 t},
~~(n^6+in^7) \mapsto (n^6+in^7)e^{i(\omega_1-\omega_2) t},
\label{rotcomp}
\ee
while the components $\{n^3,n^8\}$, which lie in the Cartan subalgebra of $su(3)$, remain unaffected.
The corresponding Noether currents are given by
\be
\begin{split}
j_\mu^{(\varphi)} &= -n^1\partial_\mu n^2 + n^2 \partial_\mu n^1 -n^6\partial_\mu n^7 + n^7 \partial_\mu n^6\, ,
\\
j_\mu^{(\psi)} &=-n^4\partial_\mu n^5 + n^5 \partial_\mu n^4 +n^6\partial_\mu n^7 - n^7 \partial_\mu n^6\, .
\end{split}
\label{Noether-j}
\ee
They yield two conserved Noether charges,
\be
\begin{split}
Q_\varphi &= \int d^2 x j_0^{(\varphi)} =
2\pi \int rdr \left\{\omega_1 \sin^2(2F)\cos^2 G
+ (\omega_1-\omega_2)\sin^4 F \sin^2(2G) \right\}  \, , \\
Q_\psi &= \int d^2 x j_0^{(\psi)} = 2\pi \int rdr \left\{
\omega_2 \sin^2(2F)\sin^2 G - (\omega_1-\omega_2)\sin^4 F \sin^2(2G)
 \right\} \, ,
\end{split}
\label{2Q}
\ee
respectively.
Notably, they both include, with opposite sign, a contribution from spinning components
$\{n^6,~n^7\}$, which lie in the $SU(2)$ subgroups  associated with the composite root of $SU(3)$.

Quantities of isospining $\mathbb{C}P^2$ solitons described by the Ansatz \eqref{Z} is given as follows.
Since spacial rotations of a (2+1)-dimensional $\mathbb{C}P^2$ configuration  are generated by a single
Killing vector field $\partial /\partial \theta$, 
the corresponding
angular momentum is given by
\begin{align}
J&= \int d^2 x T_{0\theta}
\notag\\
&=4 \pi \int dr \left\{ \sin^2(2 F)[\omega_1 k_1 \cos^2 G + \omega_2 k_2 \sin^2 G ]
+(k_1-k_2)(\omega_1-\omega_2) \sin^4 F \sin^2(2G)\right\}\, \notag\\
&=2\left( k_1 Q_\varphi + k_2 Q_\psi\right)\, ,
\label{moment}
\end{align}
 where the two Noether charges \re{2Q} are defined above.
We can write the associated total energy \eqref{T00} of the configuration \re{Z} as
\begin{align}
E&= \int d^2x T_{00}
\notag\\
=& 2\pi\int rdr \left[4 \left\{ (F^\prime)^2 + (G^\prime)^2 \sin^2 F \right\} + \sin^4 F \sin^2(2G)\left\{ \frac{(k_1-k_2)^2}{r^2} + (\omega_1-\omega_2)^2 \right\} \right.
\notag \\
&\left. \qquad\qquad\qquad\qquad\quad+
\sin^2(2 F)\left\{ \cos^2 G \left( \frac{k_1^2}{r^2} + \omega_1^2\right) +\sin^2 G \left( \frac{k_2^2}{r^2} + \omega_2^2\right) \right\}
+ V\right] \, ,
\label{TotEng}
\end{align}
where the prime stands for the derivative with respect to the radial coordinate $r$.
Moreover, substituting the Ansatz \re{Z} into Eq.~\re{topZ}, one finds that the topological charge can be cast into the form
\be
Q_{\rm top}=\int d^2 x \varepsilon^{ij} \partial_i A_j = 
\left[\sin^2 F(r)\left\{k_1 \cos^2 G(r) +k_2 \sin^2 G(r)\right\}
\right]^{r\to \infty}_{r=0} \ ,
\label{topcharge2}
\ee
which is determined by the value of the profile functions at the origin and spatial infinity.

\begin{figure}[t!]
    \centering
    \includegraphics[width=1.0\linewidth]{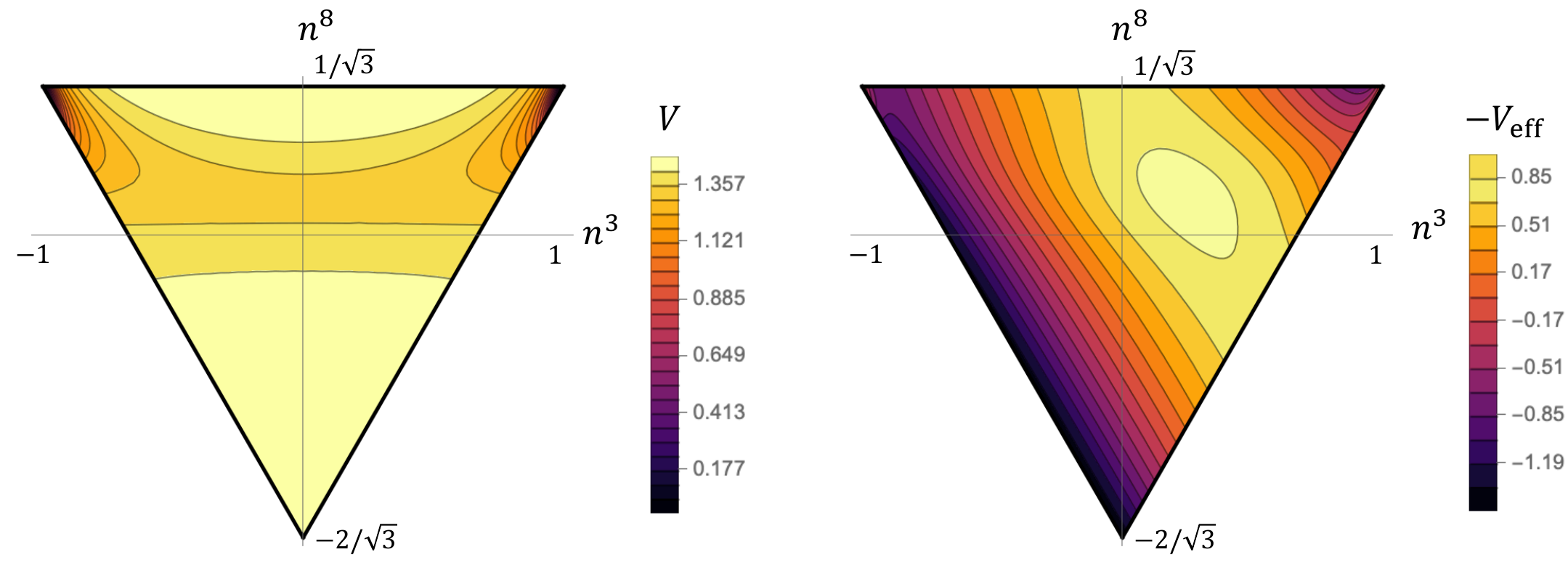}
    \caption{Value of the potential \eqref{Pot-CP2} (left) and effective potential \eqref{Veff} (right) mapped onto the toric diagram of $\mathbb{C}P^2$. For these figures, we used the parameters $\mu=1$ and $\omega_1=\omega_2=1.5$.}
    \label{fig:ToricDiagram}
\end{figure}

One can easily see that the triple vacuum potential \re{pot}, again, cannot support the existence of genuine non-embedded
$\mathbb{C}P^{2}$ isospinning solitons.
Let us now introduce a healthy potential which may support such solitons.

For the energy to be finite, at least one of vacua of the potential should make $(\partial_0 n^a)^2$ vanish. Using the Ansatz \eqref{Z}, one can obtain
\be
\begin{split}
(\partial_0 n^a)^2 &= \frac19\biggl[ \omega_1^2\left(2-3 n^3 + \sqrt3 n^8 \right)(2+3n^3 + \sqrt3 n^8)\\
&+2\omega_2^2(1-\sqrt3 n^8)\left(2+3n^3 +\sqrt3 n^8\right)
+2(\omega_1-\omega_2)^2 (1-\sqrt3 n^8)(2-3n^3+\sqrt3 n^8)
\biggr]     \, .
\end{split}
\label{iso_density}
\ee
For the use of the axial symmetric Ansatz to be reasonable, we shall employ a double vacuum potential such that $(\partial_0 n^a)^2$ becomes zero at both of the vacua.
For example, we may consider the $U(1)\times U(1)$ symmetric potential 
\be
W(n^3, n^8)=\left((2+ 3 n^3 + \sqrt 3 n^8)(2- 3 n^3 + \sqrt 3 n^8)\right)(n^8)^2 + (1-\sqrt 3 n^8) \, ,
\label{pot2}
\ee
where it has the two vacua $(n^3,n^8) = (\pm 1, 1/\sqrt{3})$.
However, the local minima of resulting effective potential $(\partial_0 n^a)^2 - W$
are not very deep, this leads to complications in the numerical analysis. Hence, we make use of the modified potential, which possesses the same vacua as Eq.~\re{pot2}, defined by
\be
V=\mu^2 \arctan (5 W(n^3, n^8)) \, .
\label{Pot-CP2}
\ee
We define the effective potential as
\begin{equation}
    V_{\rm eff}(n^3,n^8)= (\partial_0 n^a)^2 -  V \ .
    \label{Veff}
\end{equation}
In Fig.~\ref{fig:ToricDiagram}, we show the value of the potential \eqref{Pot-CP2} and effective potential \eqref{Veff} mapped on the $n^3\text{-}n^8$ plane, which is a toric diagram of $\mathbb{C}P^2$.
As one can easily observe in Fig.~\ref{fig:ToricDiagram}, the effective potential becomes non-positive in some region of the toric diagram. This is an important feature for the existence of stable soliton solutions and satisfying the viral relation
\be
    \int d^2x \left \{ (\partial_0 n^a)^2-V \right \} = \int d^2x \,  V_{\rm eff}=0.
    \label{vonLaue}
\ee
The non-trivial configurations correspond to curves on the toric diagram connecting the two vacua of the potential $(n^3,n^8) = (\pm 1, 1/\sqrt{3})$.

We impose boundary conditions that vanish the potential at both the origin and spatial infinity. We have two options:
\begin{equation}
    n^3(0)=1, ~n^8(0)=1/\sqrt 3, ~n^3(\infty)=-1, ~n^8(\infty)=1/\sqrt 3 \ ,
    \label{BC1}
\end{equation}
and
\begin{equation}
    n^3(0)=-1, ~n^8(0)=1/\sqrt 3, ~n^3(\infty)=1, ~n^8(\infty)=1/\sqrt 3 \ .
    \label{BC2}
\end{equation}
In this paper, we only consider the former \eqref{BC1} to avoid redundancy because they are physically equivalent thanks to the invariance of the Lagrangian \eqref{lag} under $n^3\to -n^3$. In terms of the profile functions, Eq.~\eqref{BC1} can be expressed as
\be
F(0)=0, ~~~G(0)=\pi/2, \quad
F(\infty)=\pi/2, ~~~G(\infty)=0 \, .
\label{BC1_FG}
\ee
Using the boundary conditions \re{BC1_FG}, we obtain
\begin{equation}
    Q_{\rm top}=k_1 \ ,
\end{equation}
and find the second integer $k_2$ remains a free non-topological parameter.
On the other hand, if we used the boundary condition $\{F(0)=G(0)=0,~
F(\infty)=G(\infty)=\pi/2\}$, which is equivalent to Eq.~\eqref{BC2}, one obtains $Q_{\rm top}=k_2$ and $k_1$ becomes a free non-topological parameter.

Our approach is based on a reformulation of the minimization problem considering the stationary points of the pseudo-energy (action) functional 
which we found numerically. This functional parametrically depends on the frequencies $\omega_1, \omega_2$, it
is constructed via the substitution of the
time-dependent Ansatz \re{Z} into the  Lagrangian \re{lagZ}.

This technique has already been applied in studies of internally rotating baby Skyrmions \cite{Halavanau:2013vsa,Battye:2013tka}, Hopfions \cite{Battye:2013xf,Harland:2013uk} and Skyrmions \cite{Battye:2014qva}.
An equivalent variational problem is to extremize the total energy functional for a fixed value of the angular momentum \cite{Ward:2003un,Battye:2005nx}. A disadvantage of this approach is that it is related to the rather complicated search for a solution of the corresponding differential-integral equation.
On the other hand, extremization of the pseudo-energy functional 
yields a system of Euler–Lagrange equations, which are analogous to the usual field equations of the  $\mathbb{C}P^2$  model.

\begin{figure}[h!]
\begin{center}
\setlength{\unitlength}{0.1cm}
\includegraphics[height=.28\textheight, angle =0]{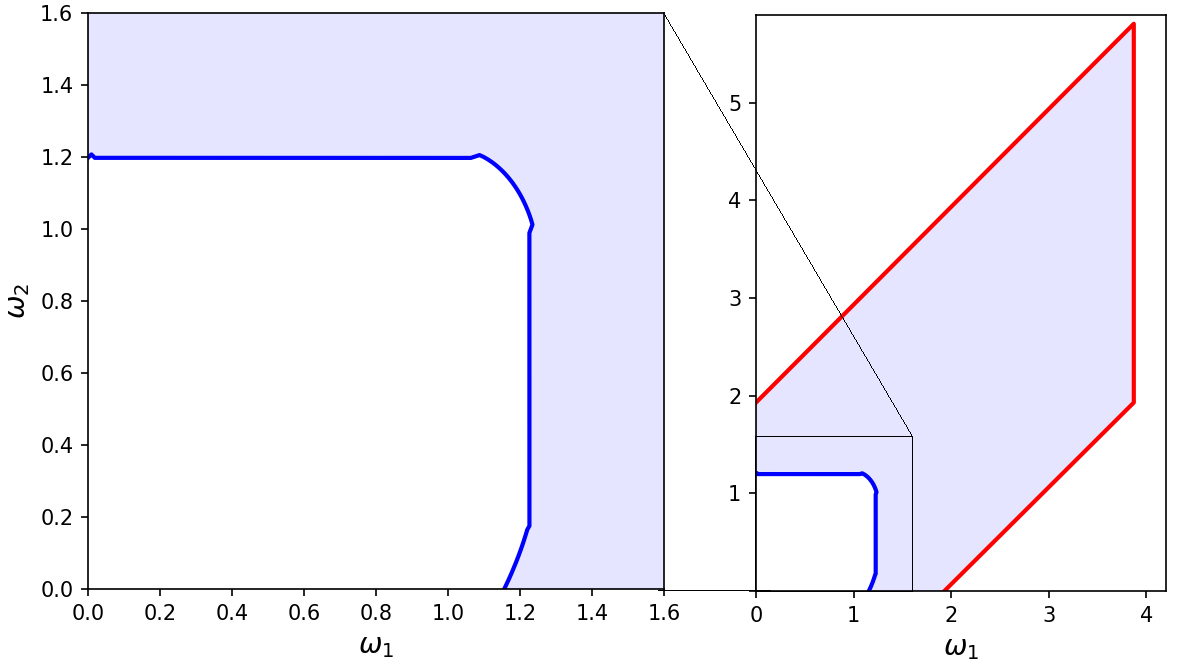}
\end{center}
\caption{Right plot: estimated lower and upper \re{domain} frequency bounds, in blue and in red respectively, with a domain of permitted frequencies in purple. Left plot: more detailed area of excluded frequencies \re{vonLaue}.}
\label{fig_freqs}
\end{figure}

Considering the asymptotic expansion of the fields around the vacuum
$F(r)\approx \pi/2 - f(r),~~G(r)\approx g(r)$, where $f(r),g(r)$ are perturbative excitations of the components,
we obtain corresponding linearized equations for the fluctuations of the fields,
\be
f^{\prime\prime} + \frac{f^\prime}{r} - ( 15 \mu^2 -\omega_1^2  )f =0\, ,\qquad
g^{\prime\prime} + \frac{g^\prime}{r} - \biggl(\frac{15}{4}\mu^2 - (\omega_1 - \omega_2)^2\biggr)g =0\,\, .
\label{linear}
\ee
Therefore, the effective mass of the excitations of the
component $F$ is $m_f=\sqrt{15\mu^2 - \omega_1^2}$ while the effective mass
of the excitations of the
component $G$ is $m_g=\sqrt{\frac{15}{4}\mu^2 - (\omega_1-\omega_2)^2}$. In other words, localized
configurations with exponentially decaying tail may exist if
\begin{equation}
\begin{aligned}
\left\{\begin{matrix}|\omega _{1}| \leq \sqrt{15} \mu \\
|\omega_1-\omega _{2}| \leq \frac{\sqrt{15}}{2}\mu\, .
\\\end{matrix}
\right.
\end{aligned}
\label{domain}
\end{equation}
There also exists ``lower'' frequencies bound which can be evaluated from the
virial relation \re{vonLaue}. 
The virial relation requires that the effective potential must possess at least one nodal point to support stable soliton solutions.
The lower frequency bound is related to the violation of this condition. For the potential \re{Pot-CP2}, numerical estimations of the lower critical frequencies are shown in  Fig.~\ref{fig_freqs}, left plot.

\section{Numerical results}\label{sec:results}

Stationary isospinning $\mathbb{C}P^2$ solutions for given winding numbers $\{ k_1,k_2\}$
are constructed by minimizing the energy functional \re{TotEng} with the potential \eqref{Pot-CP2} for fixed angular frequencies $\omega_1$ and $\omega_2$.
To find numerical solutions of the corresponding two coupled reduced second-order ordinary differential equations, which arise after the substitution of the Ansatz \re{Z} into the Euler-Lagrange equations,
we used the software package
CADSOL based on the Newton-Raphson algorithm \cite{schoen}.

\begin{figure}[h!]
\begin{center}
\setlength{\unitlength}{0.1cm}
\includegraphics[height=.34\textheight, angle =-90]{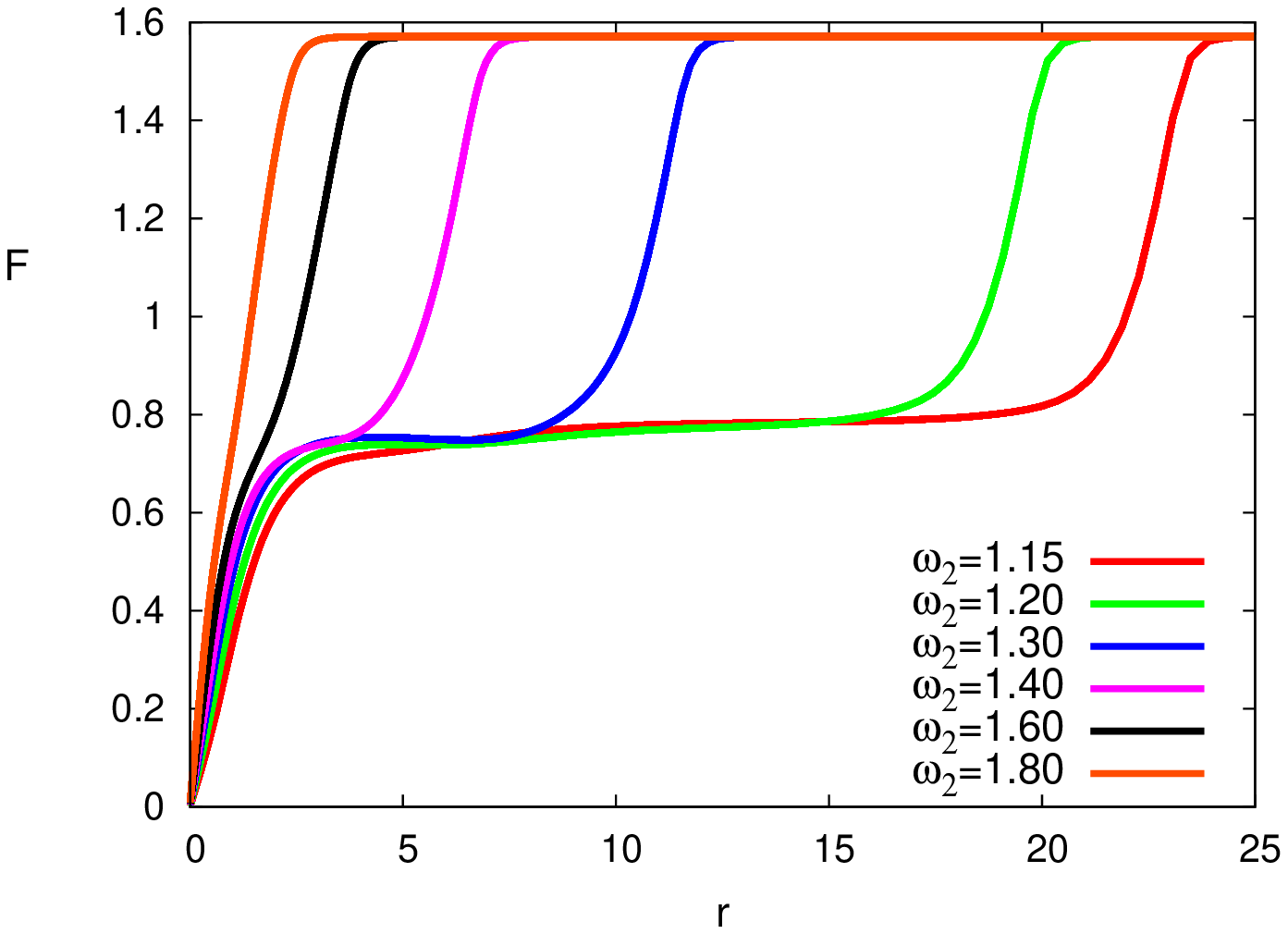}
\includegraphics[height=.34\textheight, angle =-90]{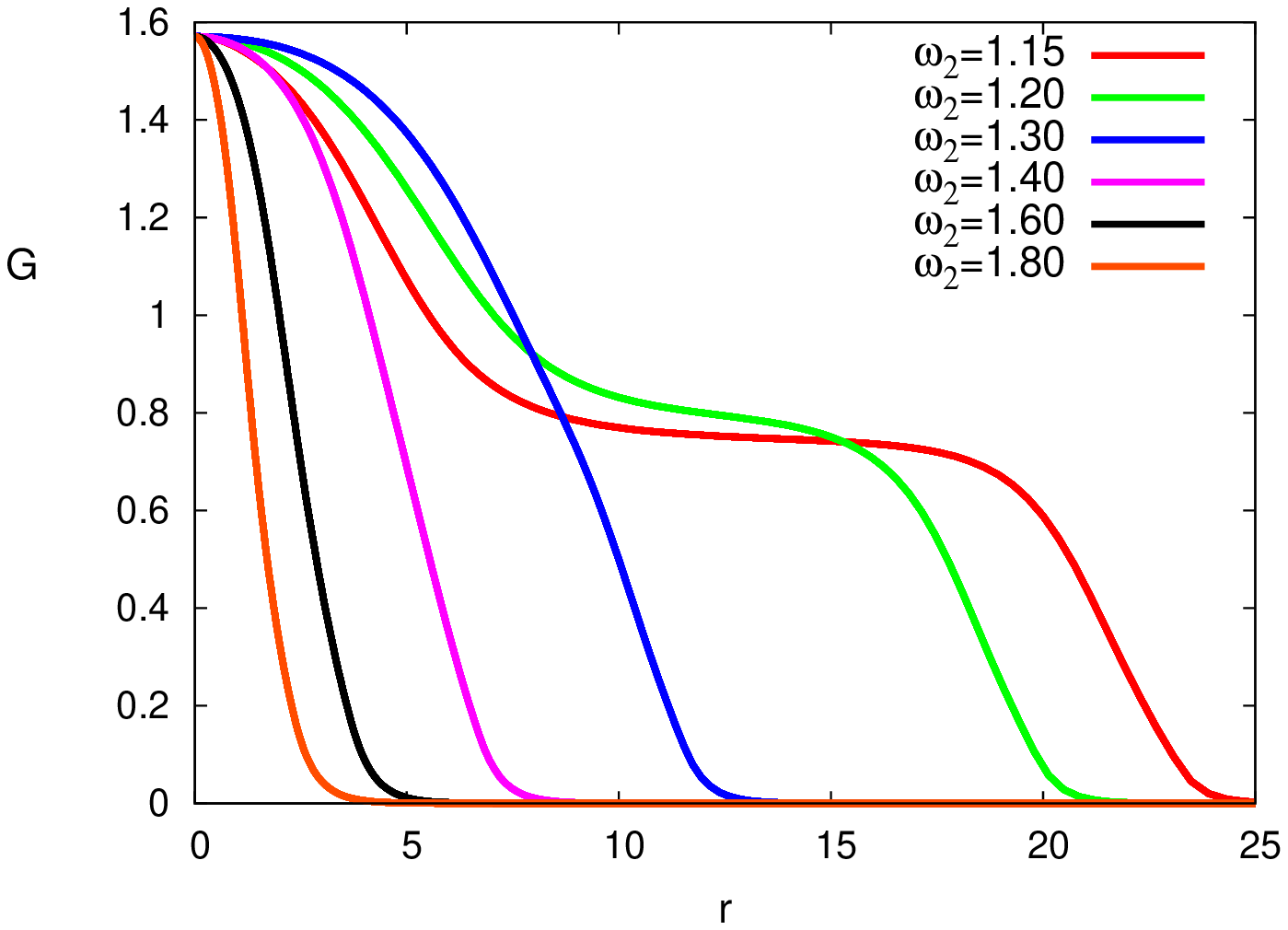}
\includegraphics[height=.34\textheight, angle =-90]{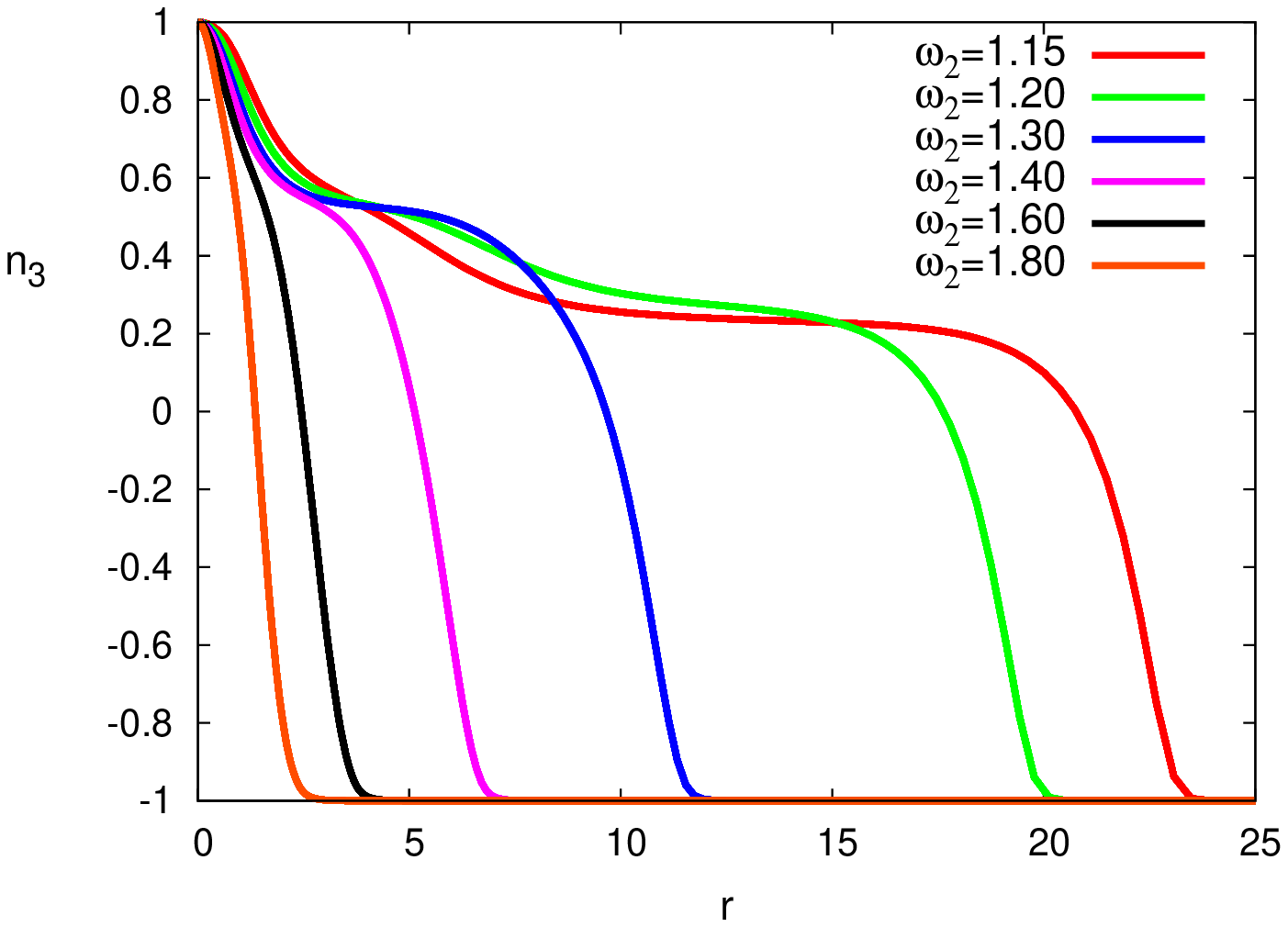}
\includegraphics[height=.34\textheight, angle =-90]{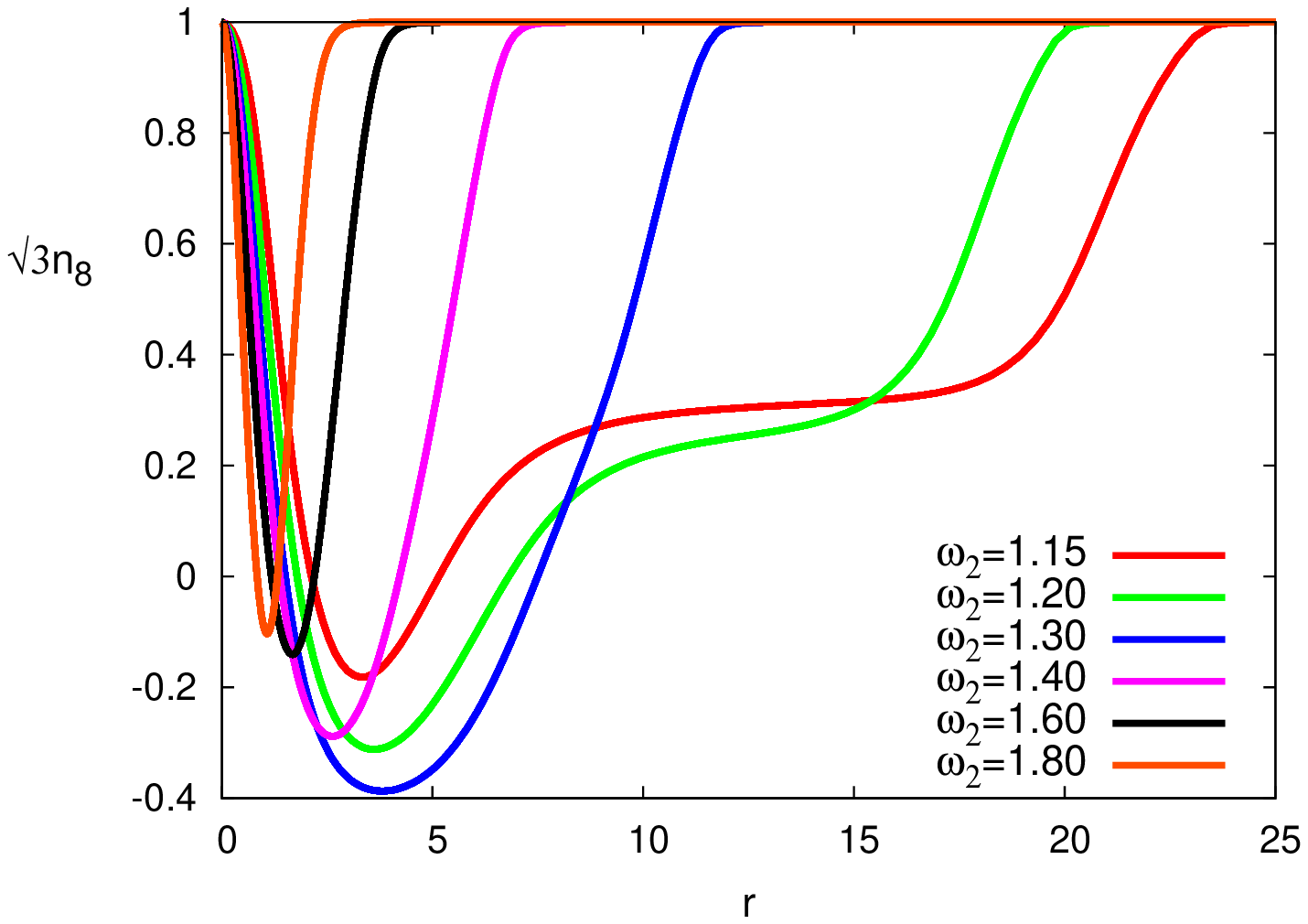}
\end{center}
\caption{The profiles of the field components $\{F,G\}$ of the  $k_1=3,k_2=1$ isospinning $\mathbb{C}P^2$ solitons  (upper plots) and the field components $(n^3, \sqrt 3 n^8)$ (bottom plots)
are plotted as functions of the radial variable $r$ at
$\omega_1=1.20$ for some set of values
of the second angular frequency $\omega_2$ at $\mu^2=1$. }
\label{fig1}
\end{figure}

We impose the above-mentioned vacuum boundary conditions \eqref{BC1_FG} on the profile function $F(r)$ and $G(r)$,
and have made use of a sixth-order finite difference scheme, where the
system of equations is discretized on a grid with a typical size of 529 points in the radial direction.
In our numerical scheme, we map the infinite interval of the variable $r$ onto the compact radial coordinate
$\tilde{r}=\frac{r/r_0}{1+r/r_0}  \in [0:1]$. Here, $r_0$ is a real scaling constant, which is used to improve the accuracy of the numerical solution.
Typically, it is taken as $r_0 = 2 - 6$. Estimated numerical errors are of order of $10^{-5}$. The virial identity \re{virial} is used to check the correctness of our results.

With our choice of parametrization, the input parameters are the mass parameter $\mu$,
the winding numbers $\{k_1,k_2\}$ and the angular frequencies $\{\omega_1,\omega_2\}$ restricted by the constraint \re{domain}.
For the sake of compactness, we now set $\mu=1$ and mainly consider, as a  particular example,  configurations with $k_1=3$ and $k_2=1$.

The total energy and the angular momentum of the stationary spinning solutions are given by Eqs.~\re{TotEng} and \re{moment}, respectively, and the corresponding Noether charges are defined by Eq.~\re{2Q}    accordingly. 

The localized isospinning $\mathbb{C}P^2$ soliton arises as $\omega_1$ decreases below a certain critical value, and the corresponding
total energy functional develops a local minimum. The emerging soliton is strongly localized, which means that the profile functions $F(r)$ and $G(r)$ rapidly approach the vacuum. The corresponding solutions for $\omega_1=1.2, ~ \omega_2=1.8$ are displayed in Fig.~\ref{fig1}.
\begin{figure}[h!]
\begin{center}
\setlength{\unitlength}{0.1cm}
\includegraphics[height=.34\textheight, angle =-90]{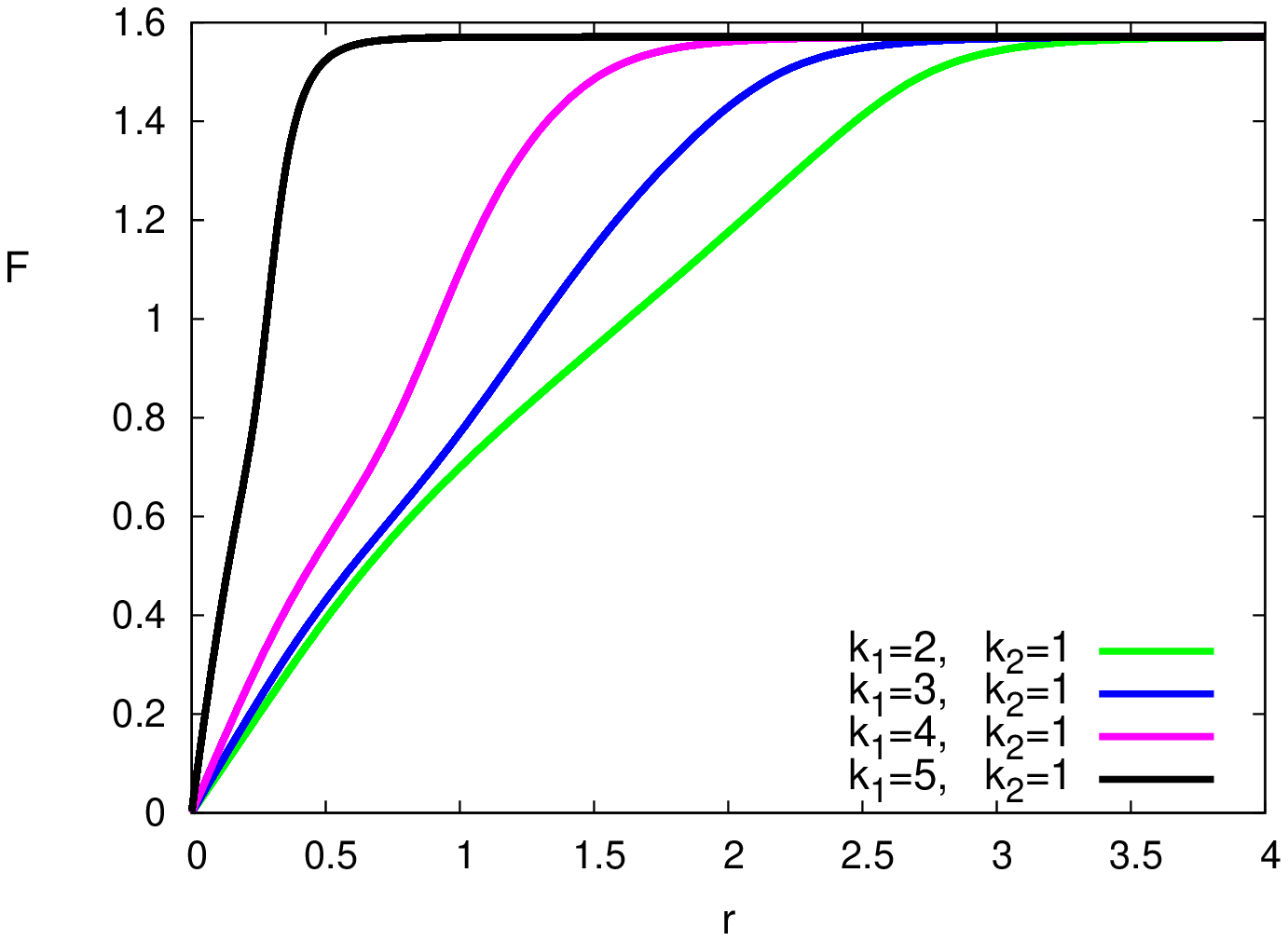}
\includegraphics[height=.34\textheight, angle =-90]{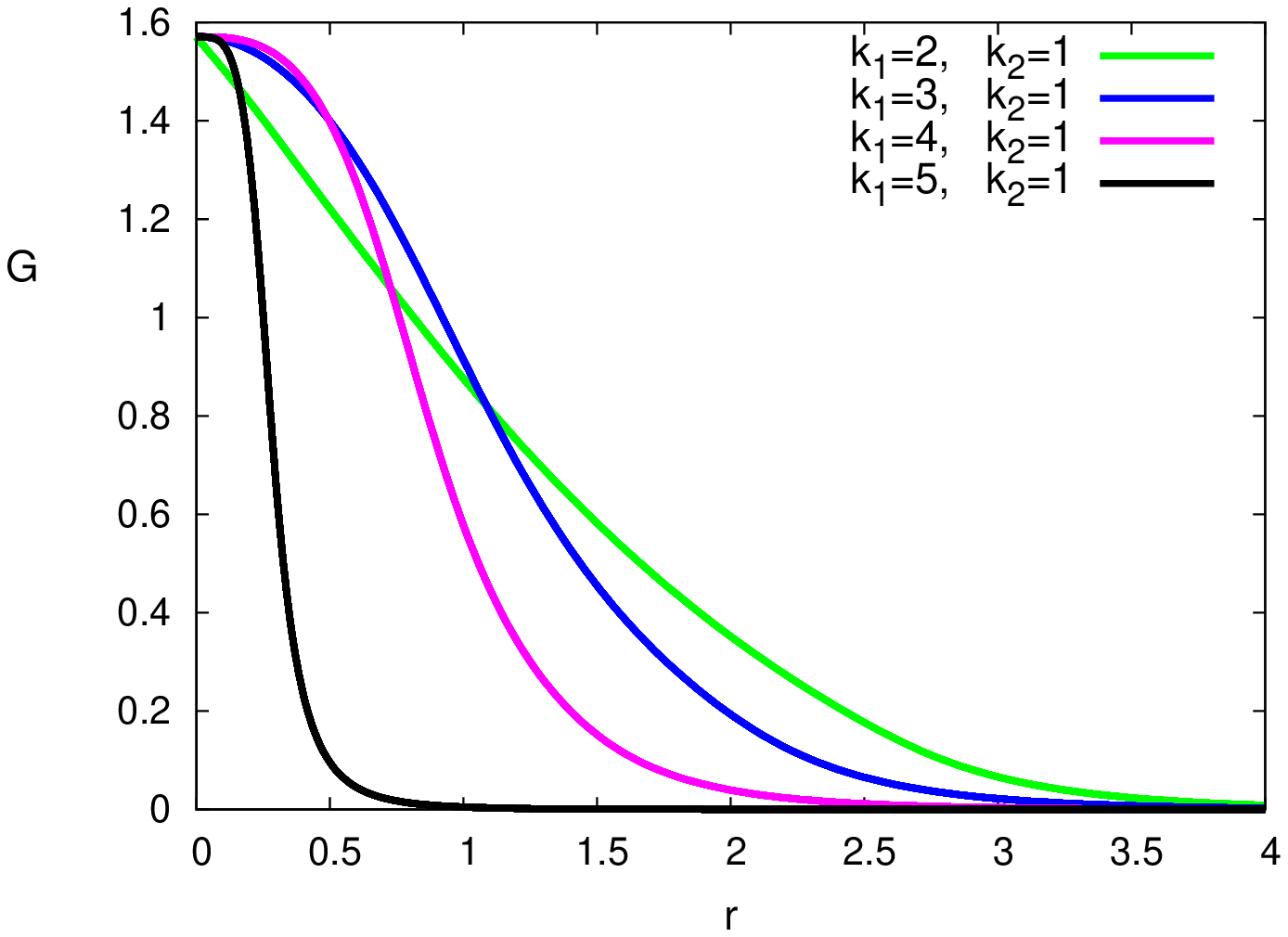}
\includegraphics[height=.34\textheight, angle =-90]{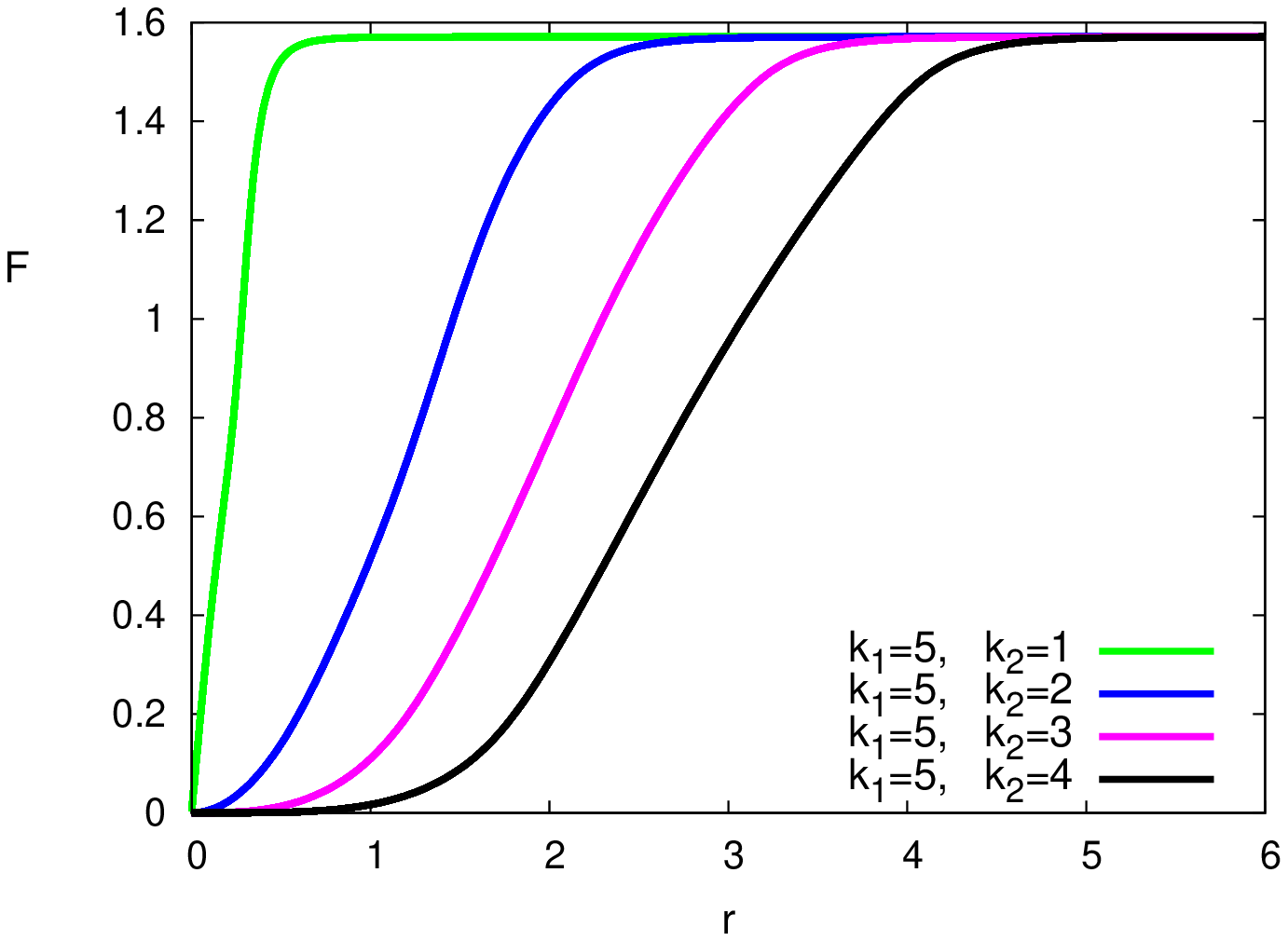}
\includegraphics[height=.34\textheight, angle =-90]{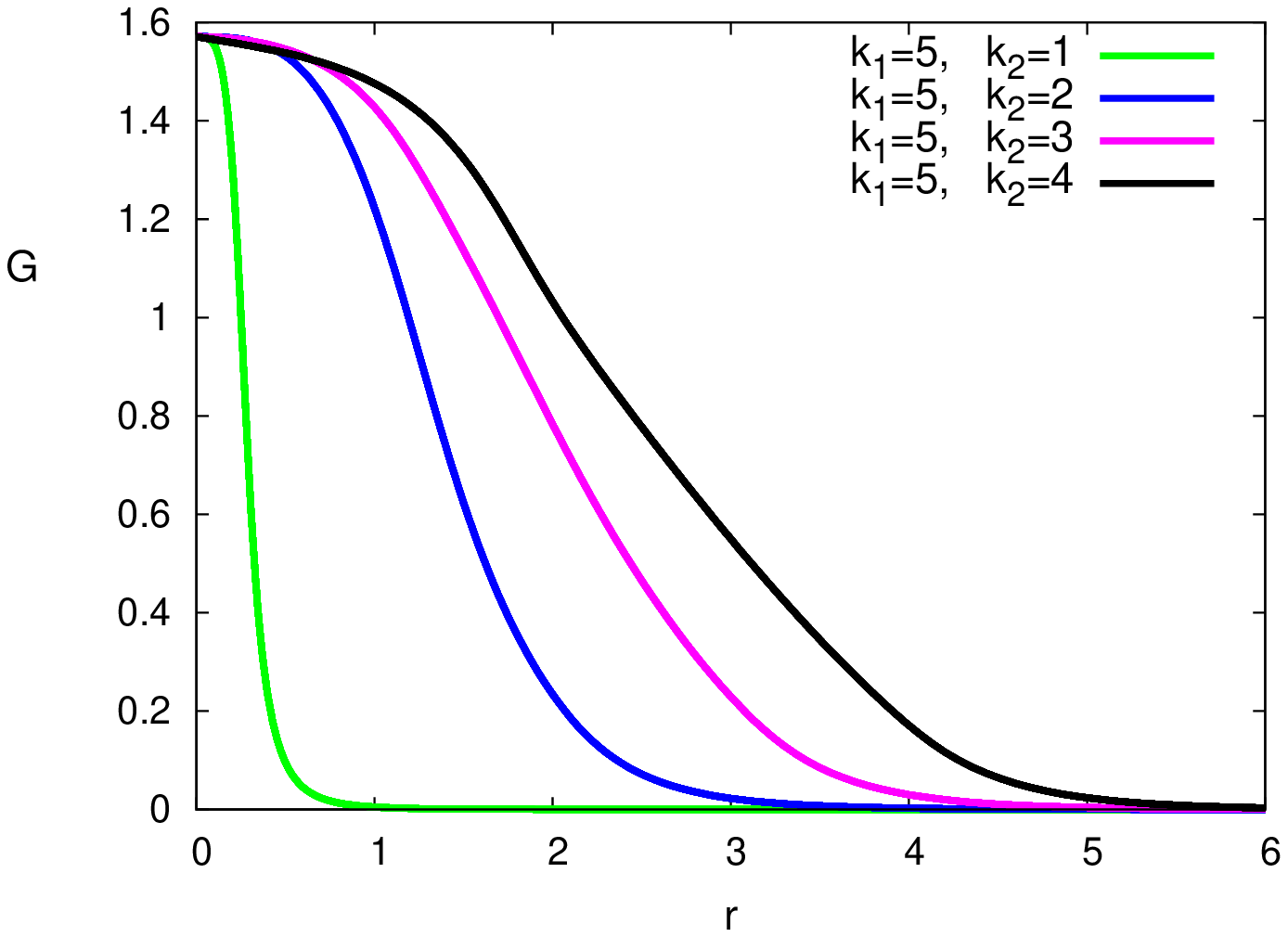}
\end{center}
\caption{The profiles of the field components $\{F,G\}$ of the  isospinning $\mathbb{C}P^2$ solitons
are plotted as functions of the radial variable $r$ at
$\omega_1=1.40,~ \omega_2=1.80$ for $2 \le k_1\le 5$ and  $k_2=1$ (upper plots) and for  $1\le k_2\le 4$ and  $k_1=5$ (bottom plots), at $\mu^2=1$. }
\label{fig0}
\end{figure}
The dependency of the solutions on the winding numbers $\{k_1,k_2\}$ reveals an interesting pattern.
Notably, we found that  $k_1>k_2$ for all solutions.
Furthermore, we did not find a solution of topological charge one, and the minimal value of the integer $k_1$ is two.

Considering the dependency of the solutions on the winding numbers, we first increase
the winding $k_1$, which is equal to the topological charge $Q_{\rm top}$ as shown in Eq.~\re{topcharge2},
keeping the second integer $k_2$ fixed. Contrary to the case of the isospinning $\mathbb{C}P^1$ solitons \cite{Ward:2003un,Mareike}, the  characteristic size of the configuration decreases as $Q_{\rm top}$
increases, see Fig.~\ref{fig0}, upper plots. Both the total energy and the angular momentum of the solitons decrease with increasing $k_1$.

On the other hand, increase of the second winding number $k_2$ for a configuration with a given topological charge $Q_{\rm top}=k_1$, expands the soliton radially outward, as displayed in Fig.~\ref{fig0}, bottom plots. The energy and the angular momentum of the configurations increase almost linearly as $k_2$ becomes larger.
\begin{figure}[h!]
\begin{center}
\setlength{\unitlength}{0.1cm}
\includegraphics[height=.34\textheight, angle =-90]{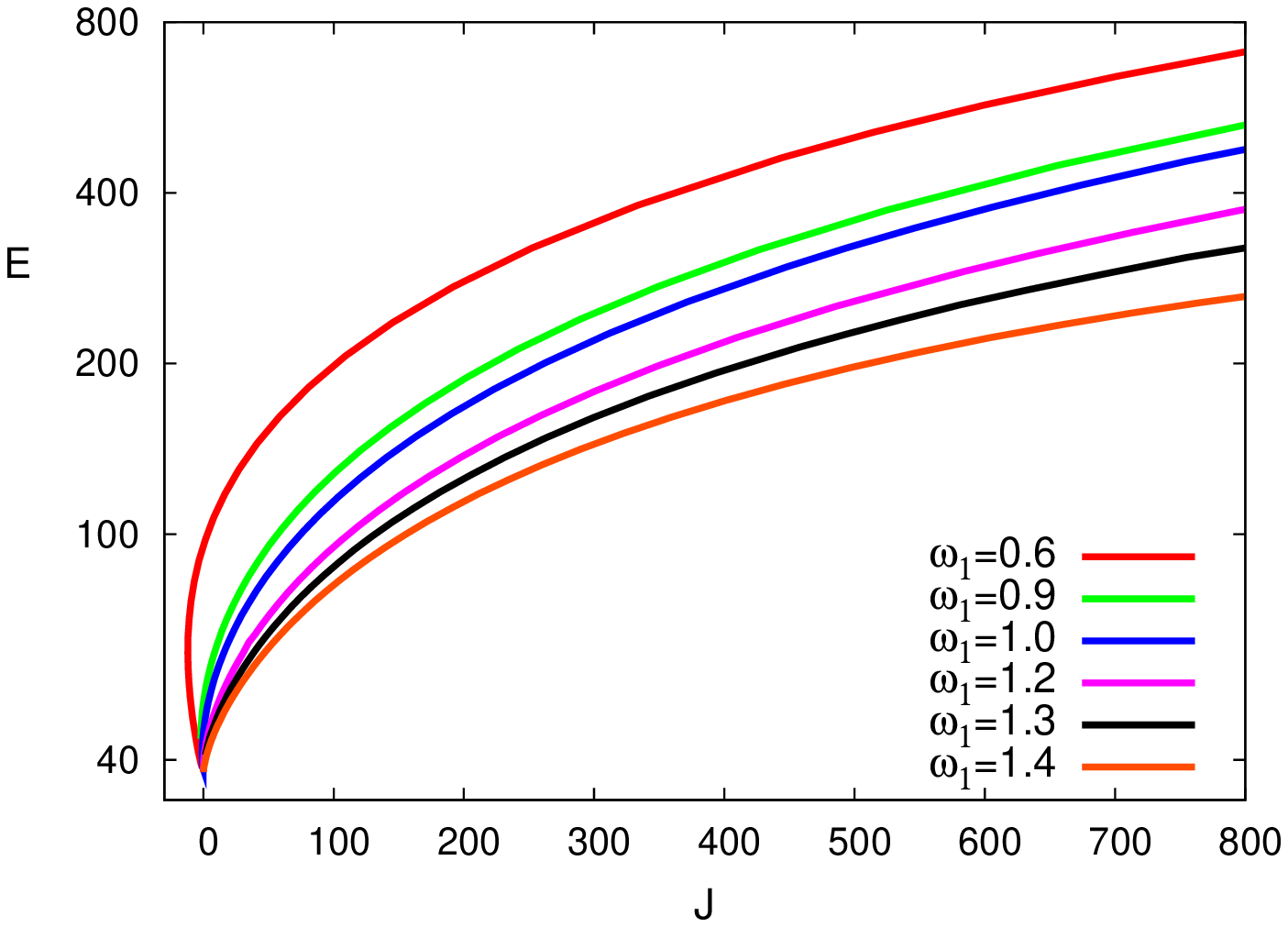}
\includegraphics[height=.34\textheight, angle =-90]{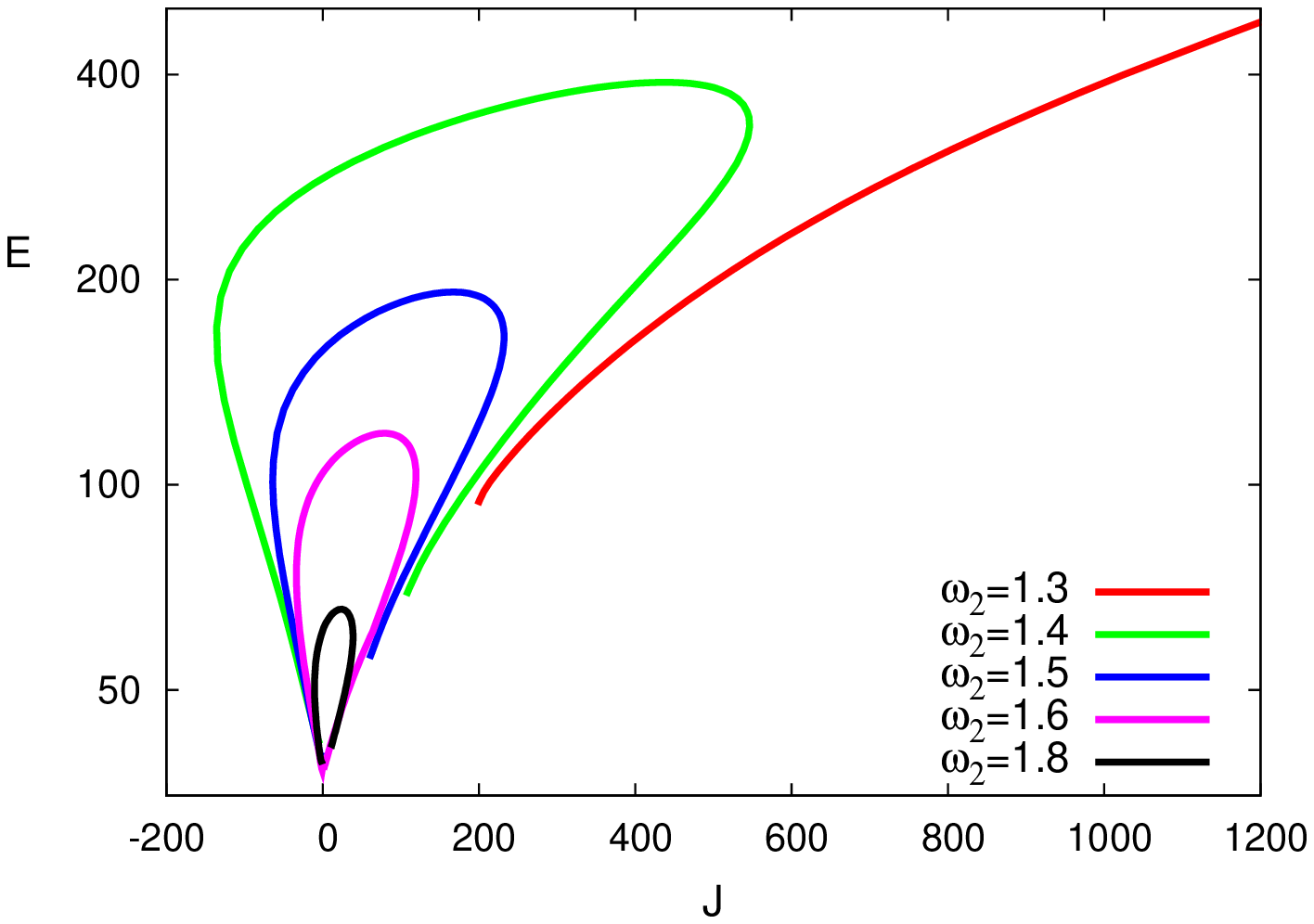}
\end{center}
\caption{The total energy $E$ \eqref{T00} of the  isospinning $\mathbb{C}P^2$ soliton with $(k_1,k_2)=(3,1)$ is plotted as a function of the momentum $J$
at $\mu^2=1$  for some set of values of the frequency $\omega_1$ (left plot) and for some set of values of $\omega_2$ (right plot). 
}
\label{fig2}
\end{figure}

We scan the parametric space
taking a particular value of the angular frequency $\omega_1$ and varying the second frequency $\omega_2$.
For a given value of the frequency $\omega_1$, both the energy and the angular momentum increase as the second frequency $\omega_2$ decreases below upper critical value, and the size of the soliton expands rapidly as the volume energy increases.  The distributions of the Noether charges also are volcano-shaped, the ``crater'' of the $Q_\psi$ 
charge distribution is more narrow, see Fig.~\ref{fig4}. Note that as the
frequency $\omega_1$ becomes relatively small, the regions of negative $Q_\varphi$ charge density may appear. Moreover the total angular momentum   $J$ can be negative, see
Fig.~\ref{fig2} which exhibits the $E(J)$ curves of the $k_1=3, k_2=1$ configurations at $\mu^2=1$. We observe that, for a given value of $\omega_1$,  increase of the angular momentum leads to a linear
relation between the energy $E$ and angular momentum $J$, see Fig.~\ref{fig2} left plot. Notably,  the slope of the lines is almost the same for all solutions in a given topological sector. We observe that the angular momentum may become negative as $\omega_1$ decreases below certain value $\lesssim 0.7$.

Variation of the second frequency $\omega_2$ yields a different picture. As $\omega_2 \gtrsim 1.33$, a branch of isospinning solitons with negative angular momentum emerges from the singular solution, see Fig.~\ref{fig2} right plot. Both the effective energy and the magnitude of the angular momentum decrease as $\omega_2$ becomes larger. Typically, there are two different configurations with the same total energy but different angular momentum and Noether charges. The branch of solutions with positive angular momentum terminates as a configuration approach the limiting embedded  $\mathbb{C}P^1$ soliton. As the angular frequency  $\omega_2$ decreases below the lower critical value $\sim1.33$, the volume energy of the soliton becomes larger than its surface energy, the configuration rapidly expands over the entire space.

\begin{figure}[h!]
\begin{center}
\setlength{\unitlength}{0.1cm}
\includegraphics[height=.21\textheight, angle =0]{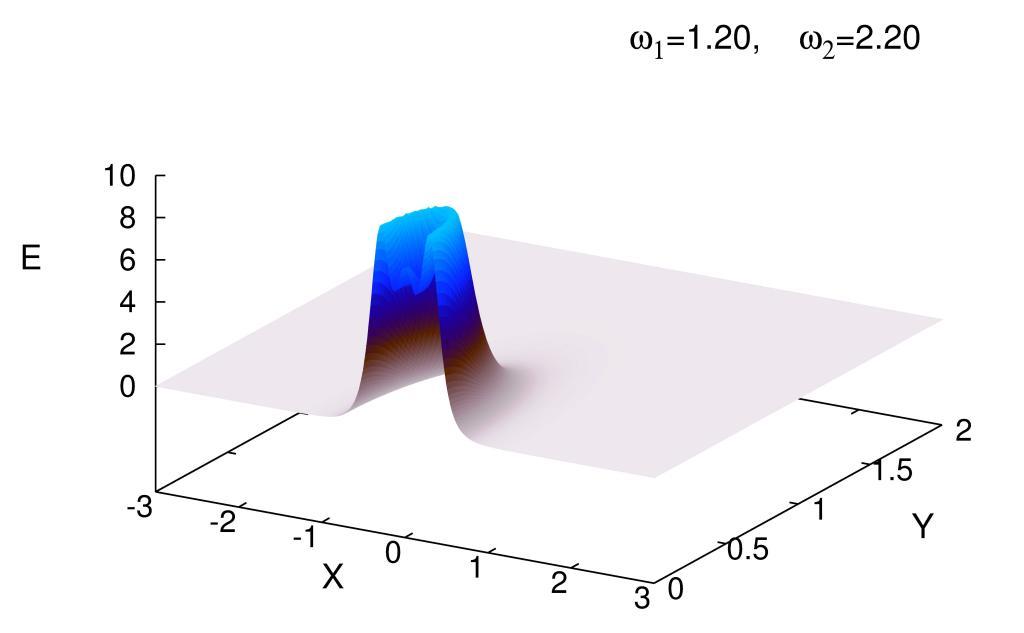}
\includegraphics[height=.21\textheight, angle =0]{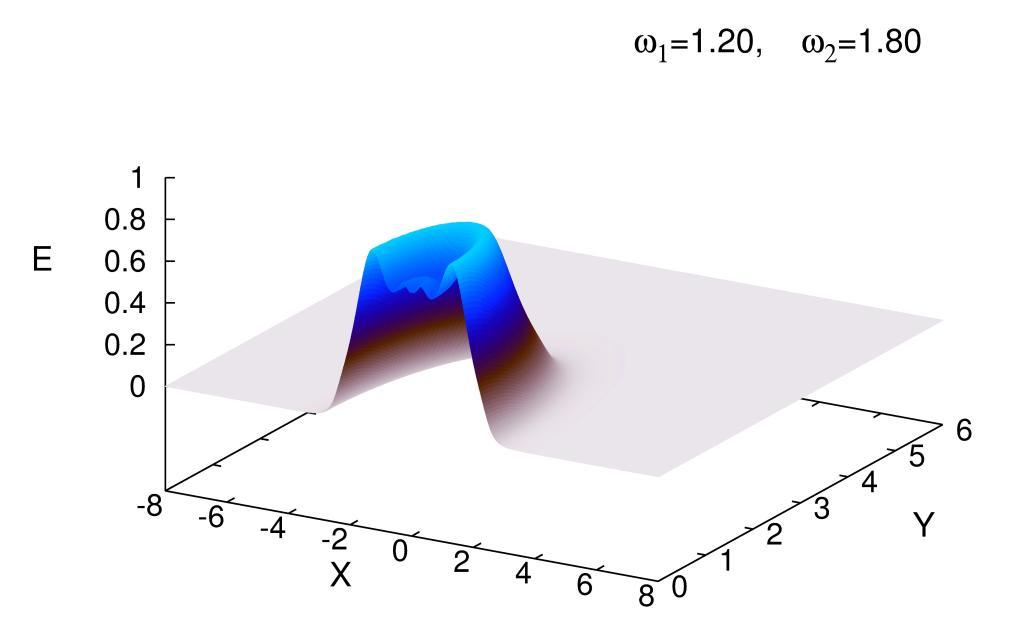}
\includegraphics[height=.21\textheight, angle =0]{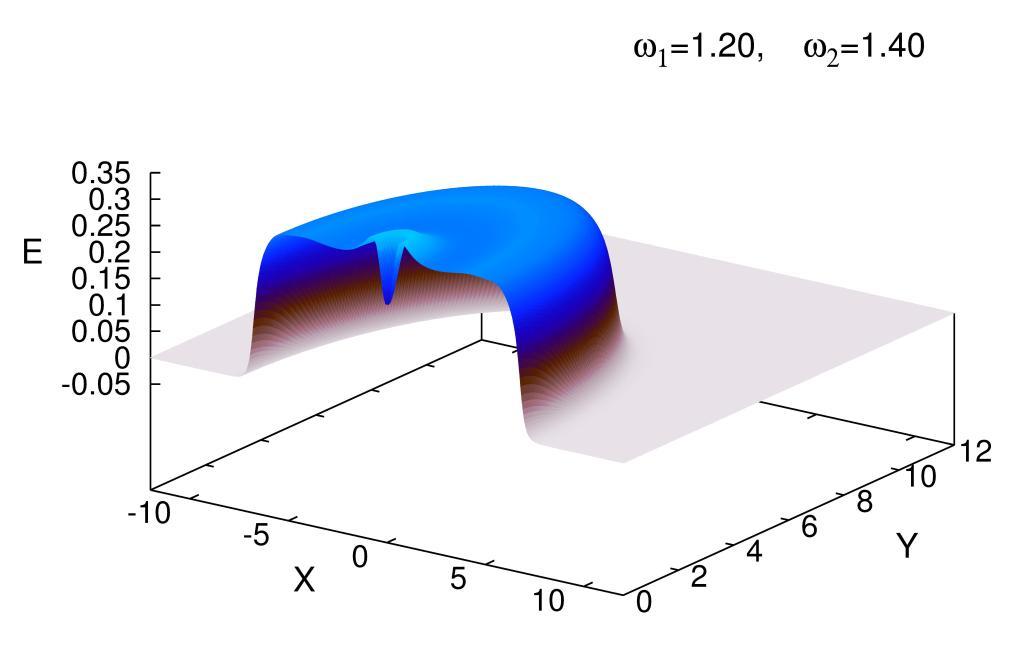}
\includegraphics[height=.21\textheight, angle =0]{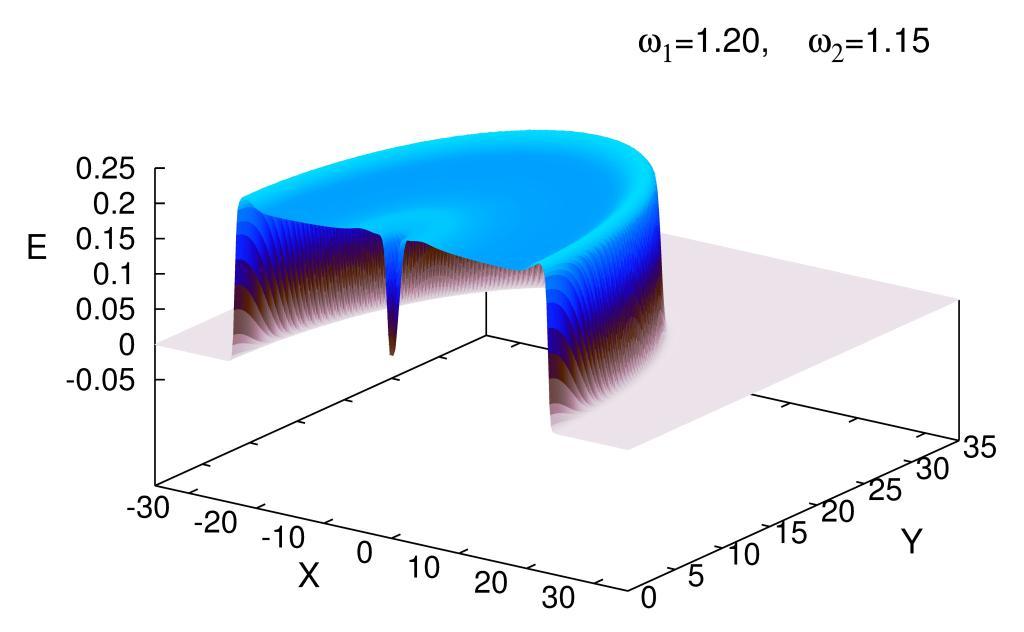}
\end{center}
\caption{The isospinning $\mathbb{C}P^2$ solitons with $ (k_1, k_2)=(3,1)$:  3d plots of the  total energy distribution versus the Cartesian coordinates $(x,y) =  (r\sin \theta, r\cos \theta)$
for some set of values
of angular frequency $\omega_2$ at $\omega_1=1.2$ and $\mu^2=1$.}
\label{fig3}
\end{figure}
\begin{figure}[h!]
\begin{center}
\setlength{\unitlength}{0.1cm}
\includegraphics[height=.20\textheight, angle =0]{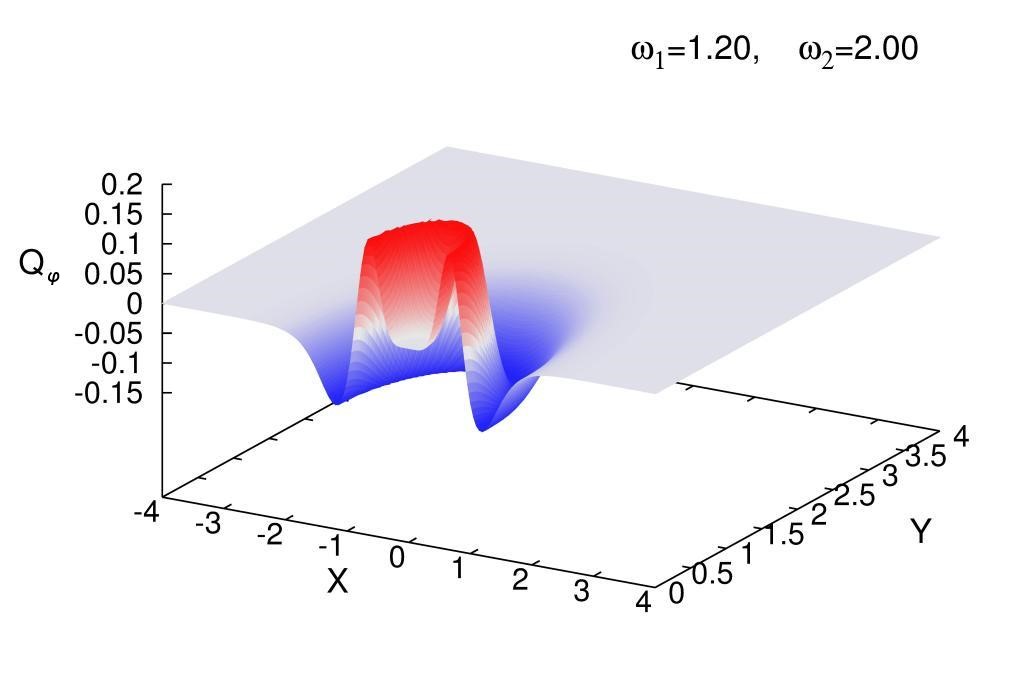}
\includegraphics[height=.20\textheight, angle =0]{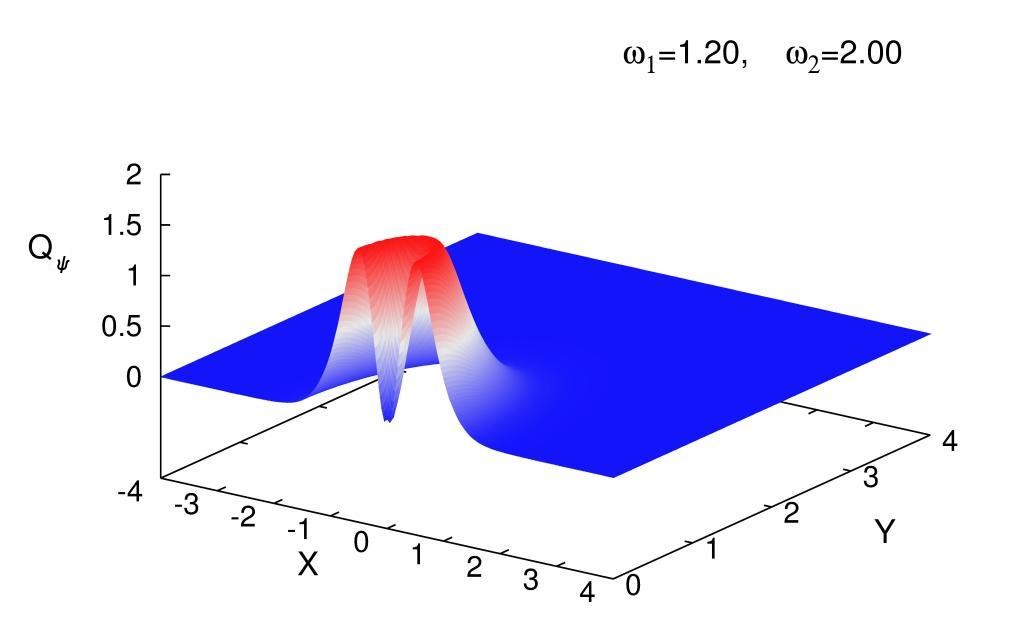}
\includegraphics[height=.20\textheight, angle =0]{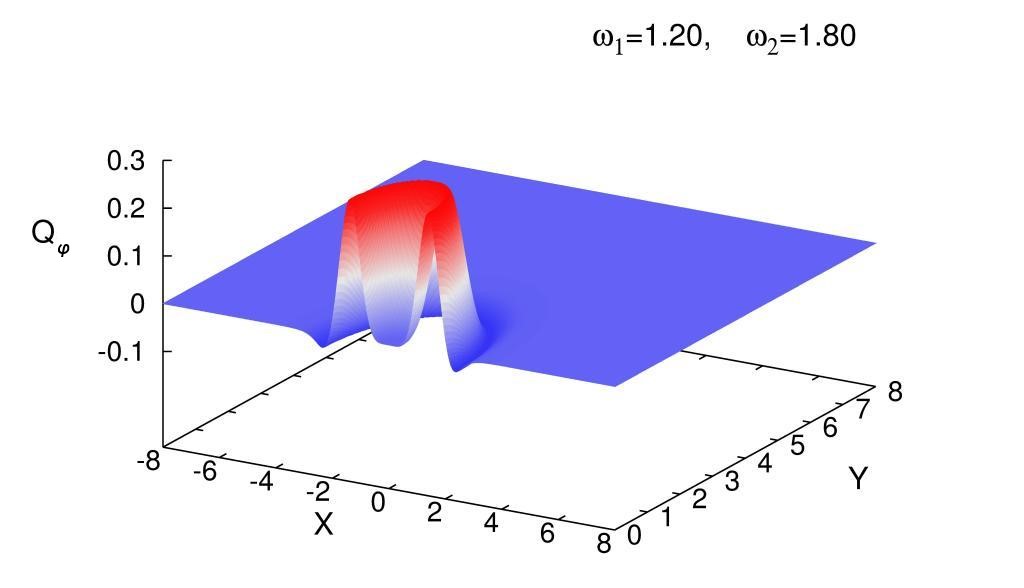}
\includegraphics[height=.20\textheight, angle =0]{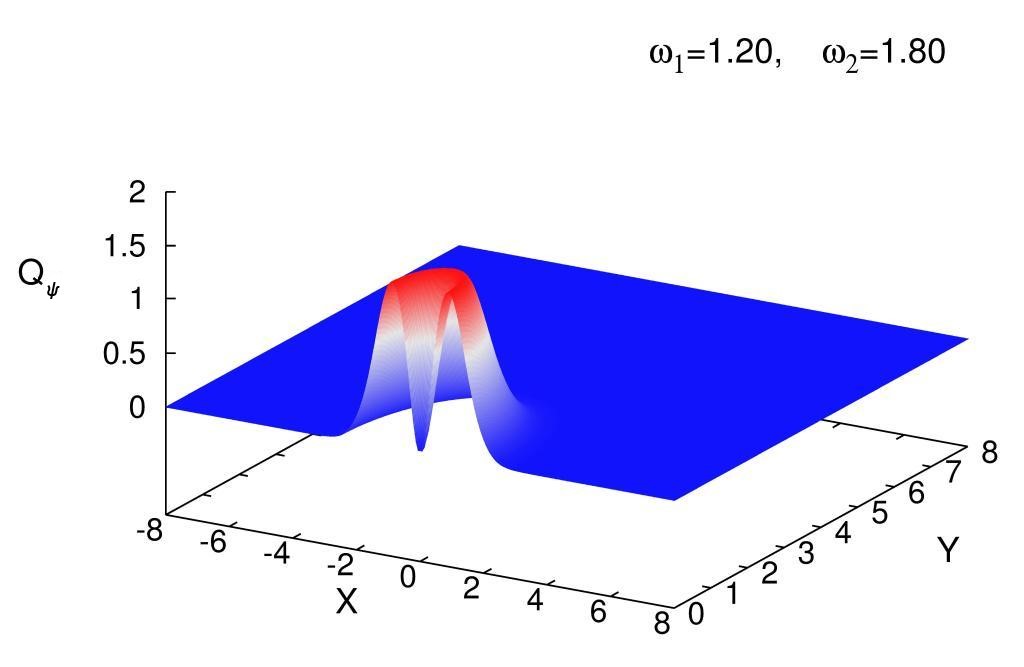}
\includegraphics[height=.20\textheight, angle =0]{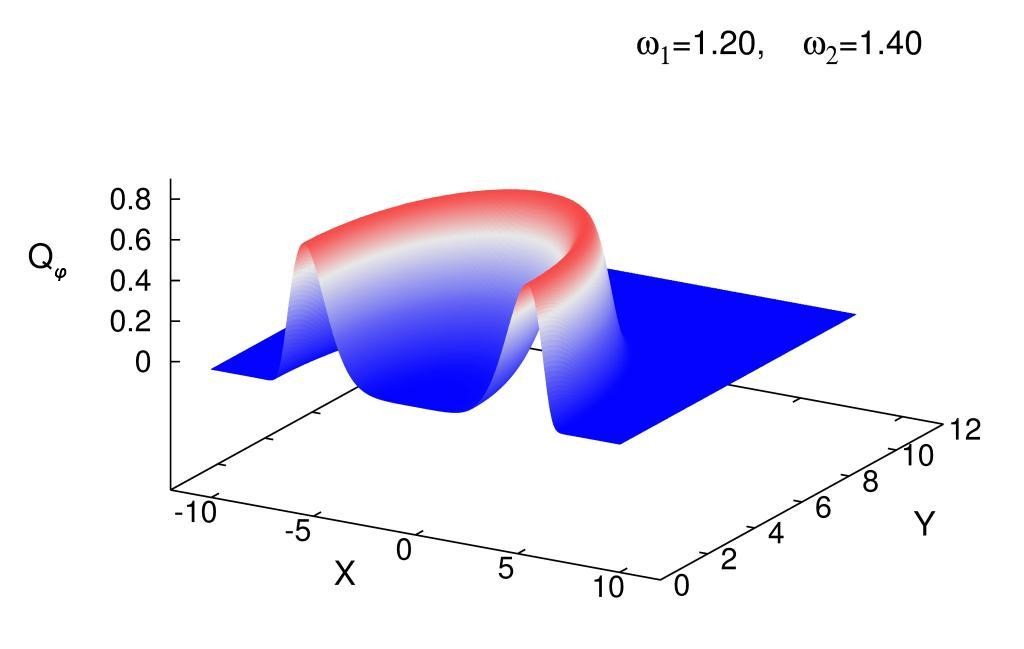}
\includegraphics[height=.20\textheight, angle =0]{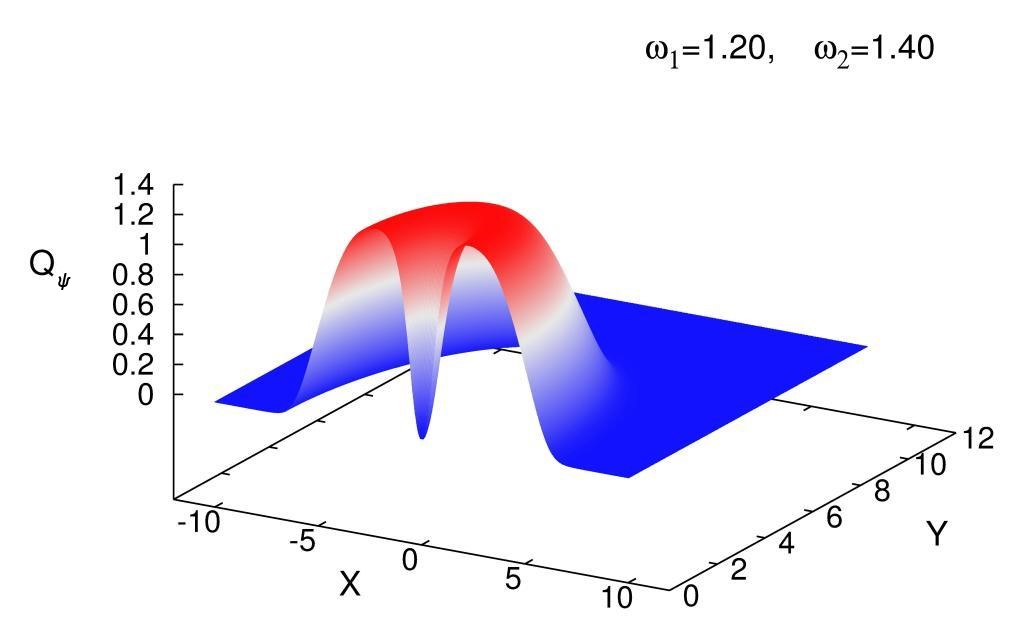}
\includegraphics[height=.20\textheight, angle =0]{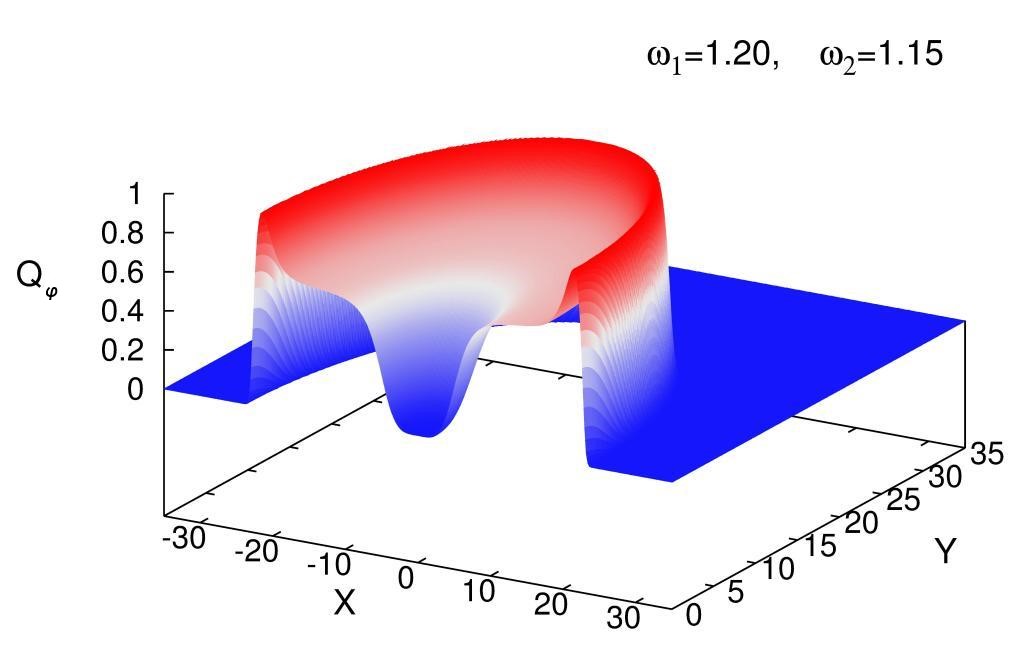}
\includegraphics[height=.20\textheight, angle =0]{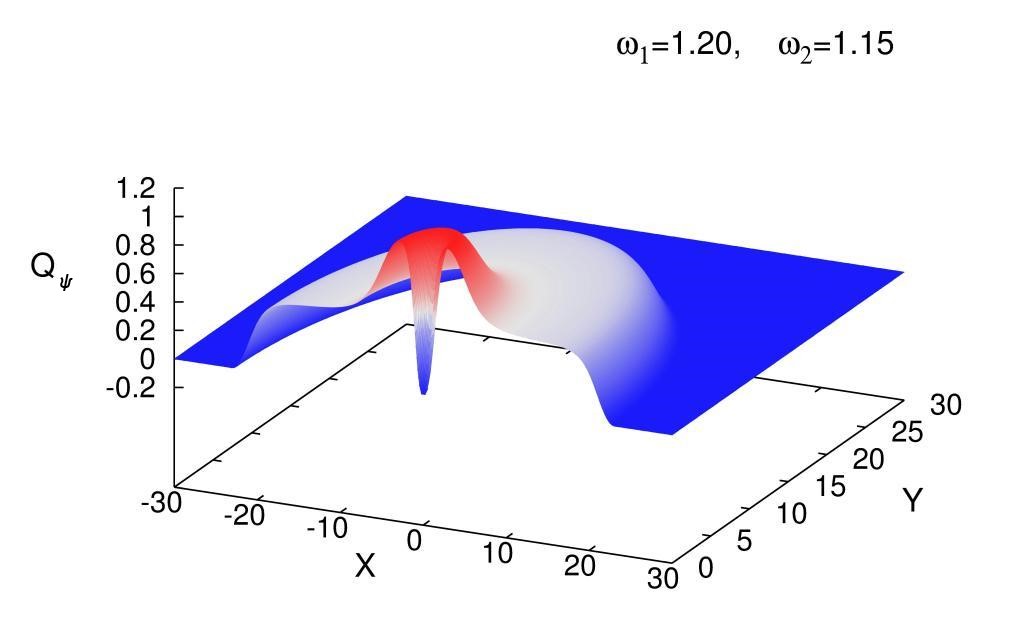}
\end{center}
\caption{The isospinning $ k_1=3, k_2=1$ $\mathbb{C}P^2$ solitons:  3d plots of the Noether charges $Q_\varphi$ (left column) and $Q_\psi$ (right column)  distributions
versus the Cartesian coordinates $x = r \sin \theta$ and $y = r \cos \theta$
for some set of values
of angular frequency $\omega_2$ at $\omega_1=1.2$ and $\mu^2=1$.}
\label{fig4}
\end{figure}
The energy density of the $k_1=3, k_2=1$ configurations forms a circular ring with a local minimum at the center, see Fig.~\ref{fig3}, upper  plots.
One can clearly distinguish two different scenarios for the evolution of the solitons. In both cases the increase of the frequency $\omega_2$ leads to
transition from the so-called
“thick-wall” limit to the
“thin-wall” limit, the configuration rapidly expands because the volume energy, together with the isorotational contribution $E_\omega$,  becomes much higher than the surface energy. The configuration
forms a compact domain with a wall that is separating the vacuum with $F=\pi/2,~ G=0$ on the exterior and confining the soliton in the interior. The profile function $F(r)$ of the compacton resembles the corresponding solution for an embedded $\mathbb{C}P^1$ soliton \cite{Ward:2003un}, in some small region around the center of the configuration  it rapidly increases from $F(0)=0$ to
$\pi/4$, then it forms an extended plateau, which ends in the surface region where the profile function approaches the vacuum, see  Fig.~\ref{fig1} left plot. As the frequency $\omega_1$ remains larger than $\sim 1.24$, the second profile function $G(r)$ monotonically decreases from $G(0)=\pi/2$ to the vacuum, as displayed in
Fig.~\ref{fig1} right plot.
This compact domain is blowing up rapidly as the angular frequency $\omega_2$ approaches its lower critical value, which corresponds to the transformation of the local minimum of the effective potential into a true vacuum. 

As the frequency $\omega_1$
approaches the critical value
$\omega_1^{cr} \sim 1.2$, the local minimum of the effective potential at the origin tends to the second vacuum, see Fig.~\ref{fig3}. The profile function $G(r)$ then also develops a plateau where $G=\pi/4$, the configuration  blows up as the second frequency decreases down to
the critical value $\omega_2\approx 1.15$.

A novelty is observed as the frequency $\omega_1$ decreases further below $\omega_1^{cr}$, the local minimum of the effective potential remains to be a false vacuum within all allowed range of values of the frequency $\omega_2$. The mass and the angular momentum of the critical solution do not diverge, they remain finite until the soliton cease to exist. The allowed range of values of the frequency $\omega_2$ is shrinking as $\omega_1 $ continues to decrease.

\section{Conclusions}\label{sec:summary}
The objective of this work has been to study
isospinning  $\mathbb{C}P^2$  soliton solutions
in 2+1 dimensions. The model is stabilized by additional coupling to a potential (mass) term by analogy with the similar consideration of the internally rotating $O(3)$ solitons \cite{Leese:1991hr,Abraham:1991ki,Ward:2003un,Mareike}.   For constructing non-embedded soliton solutions of the model, we have made use of the $U(1)\times U(1)$ invariant parametrization with two profile functions and two angular frequencies $\omega_1, \omega_2$.
We have found that stable isospinning   $\mathbb{C}P^2$ solitons exist within a certain range of the two frequencies. The upper bound \re{domain} is related with the condition of the existence of exponentially localized solutions. The lower critical frequencies are related with minima of the potential function \re{Pot-CP2}.
Interestingly, we have not found isospinning  $\mathbb{C}P^2$ solitons with topological charge one, moreover, all solutions we found are restricted by the condition on the winding numbers $k_1 > k_2$.
Our calculations have shown that, in general, the pattern of dynamical evolution of the isospinning   $\mathbb{C}P^2$ configurations is quite different from the corresponding case of the non-linear $O(3)$ sigma model. While in the latter case there is a simple linear relation between the azimuthal winding number and the topological charge of the configuration, the topological charge of the  $\mathbb{C}P^2$ soliton depends only on the first winding number $k_1$, while the second one remains a free parameter of the model.
Thus, the relation between the total energy and angular momentum of the configuration is not necessarily linear.
Considering limiting behavior of the solutions, we have found two different scenarios. In the first case a configuration approaches the limiting embedded   $\mathbb{C}P^1$ solution, the second situation corresponds to the unlimited expansion of the soliton.

Now, let us discuss possible future directions.
In this paper, we have discussed only rotationally invariant configurations parameterized by the Ansatz \re{Z}. It will be natural to investigate if, by analogy with the baby Skyrmions \cite{Halavanau:2013vsa,Battye:2013tka,Mareike}, the rotational symmetry of the isospinning  $\mathbb{C}P^2$ multisolitons
can be violated and the configuration
may become unstable with respect to decay into partons at some critical values of the
angular frequencies. One can expect the critical behavior of isospinning configurations strongly depends on the structure of the potential term, it would be interesting to study how our
results change for another choice of the potential.

Note that in a recent publication \cite{Amari:2024adu} a possibility of stabilizing topological solitons in a gauged $O(3)$-sigma model was studied. Since the internal rotations of the scalar field can be promoted to the time-dependent gauge transformations,
one of the directions in which the present work can be further continued, is to study
soliton solutions of the gauged $\mathbb{C}P^N$ model.  Finally, it might also be interesting to consider isospinning solitons of the $\mathbb{C}P^2$ model with a Skyrme-like term,
and
an extension to composite solitons
composed of different kinds
\cite{Eto:2005sw,Eto:2006pg}
such as Q-lump strings ending on
a domain wall \cite{Gauntlett:2000de,Isozumi:2004vg,Eto:2008mf}.
We hope that we can address these issues in our future work.

\section*{Acknowledgment}
Y.S. would like to thank the Hanse-Wissenschaftskolleg Delmenhorst for support and  hospitality.
This work is supported in part by JSPS KAKENHI [Grants No. JP23KJ1881 (YA) and No. JP22H01221 (MN)] and the WPI program ``Sustainability with Knotted Chiral Meta Matter (SKCM$^2$)'' at Hiroshima University.


 \end{document}